\def\year{2019}\relax
\documentclass[letterpaper]{article} 
\usepackage{aaai19}  
\usepackage{times}  
\usepackage{helvet}  
\usepackage{courier}  
\usepackage{url}  
\usepackage{graphicx}  
\usepackage[ruled,linesnumbered]{algorithm2e}
\usepackage{subcaption}
\usepackage{amsmath, amsthm}
\usepackage{multirow}

\newtheorem{theorem}{Theorem}
\newtheorem{lemma}{Lemma}

\newtheorem{definition}{Definition}

\begin{document}
%
\title{Communication-Optimal Distributed Dynamic Graph Clustering}

\author{Chun Jiang Zhu,\textsuperscript{1} Tan Zhu,\textsuperscript{1} Kam-Yiu Lam,\textsuperscript{2} Song Han,\textsuperscript{1} Jinbo Bi\textsuperscript{1}\\
\textsuperscript{1} Department of Computer Science and Engineering, University of Connecticut, Storrs, CT, USA\\
\{chunjiang.zhu, tan.zhu, song.han, jinbo.bi\}@uconn.edu\\
\textsuperscript{2} Department of Computer Science, City University of Hong Kong, Hong Kong, PRC\\
cskylam@cityu.edu.hk
}
\maketitle
\begin{abstract}
We consider the problem of clustering graph nodes over large-scale dynamic graphs, such as citation networks, images and web networks, when graph updates such as node/edge insertions/deletions are observed distributively. We propose communication-efficient algorithms for two well-established communication models namely the message passing and the blackboard models. Given a graph with $n$ nodes that is observed at $s$ remote sites over time $[1,t]$, the two proposed algorithms have communication costs $\tilde{O}(ns)$ and $\tilde{O}(n+s)$ ($\tilde{O}$ hides a polylogarithmic factor), almost matching their lower bounds, $\Omega(ns)$ and $\Omega(n+s)$, respectively, in the message passing and the blackboard models. More importantly, we prove that at each time point in $[1,t]$ our algorithms generate clustering quality nearly as good as that of centralizing all updates up to that time and then applying a standard centralized clustering algorithm. We conducted extensive experiments on both synthetic and real-life datasets which confirmed the communication efficiency of our approach over baseline algorithms while achieving comparable clustering results.
\end{abstract}

\section{Introduction}

Graph clustering is one of the most fundamental tasks in artificial intelligence and machine learning \cite{GMT+14,TGC+14,ALL+16}.
Given a graph consisting of a node set and an edge set, graph clustering asks to partition graph nodes into clusters such that nodes within the same cluster are ``densely-connected''  by graph edges, while nodes in different clusters are ``loosely-connected''.
Graph clustering on modern large-scale graphs imposes high computational and storage requirements, which are too expensive, if not impossible, to obtain from a single machine.
In contrast, distributed computing clusters and server storages are a popular and cheap way to meet the requirements.
Distributed graph clustering has received considerable research interests \cite{HYCC07,YX15,CSW+16,SZ17}.
However, the dynamic nature of modern graphs makes the clustering problem even more challenging.
We discuss several motivational examples and their characteristics as follows.

\begin{figure}[t]
\center
\includegraphics[width=0.8\columnwidth]{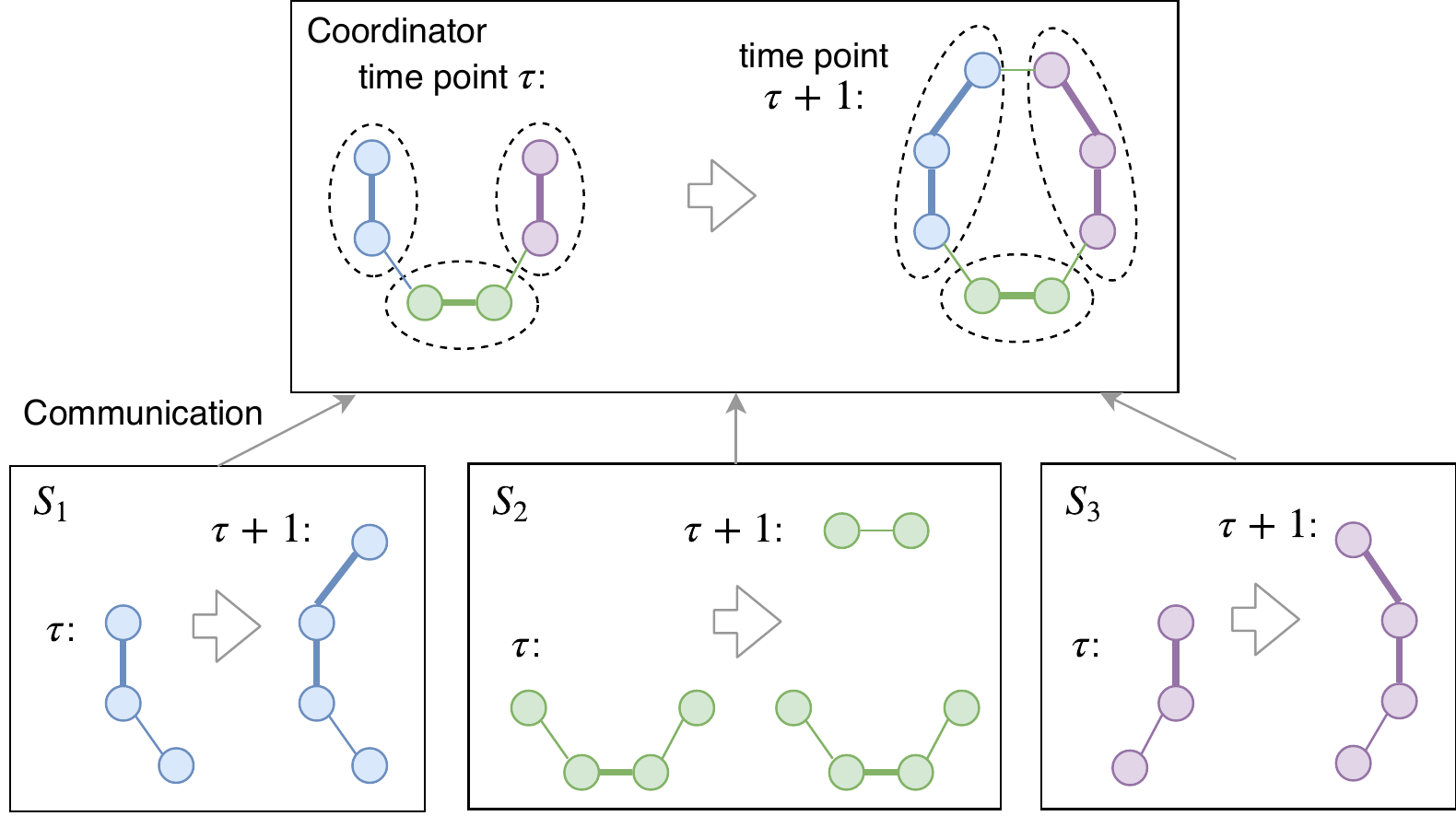}
\caption{\small Illustration of distributed dynamic graph clustering. Thick edges have an edge weight 3 while thin edges have an edge weight 1. Clustering results are evolving over time.}
\label{fig:illustration}
\vspace{-0.25in}
\end{figure}

{\bf \noindent Citation Networks.}
Graph clustering on citation networks aims to generate groups of papers/manuscripts/patents with many similar citations.
This implies that the authors within each cluster share similar research interests.
The clustering results can be useful for recommending research collaboration, e.g. in ResearchGate.
Large-scale citation networks, e.g. the US patent citation network (1963-1999)\footnote{https://snap.stanford.edu/data/cit-Patents.html}, contain millions of patents and tens of millions of citations, and they are dynamic with frequent insertions. New papers are published everyday with new citations to be added to the network graph.
Citation networks usually have negligible deletions because very few works get revoked.

{\bf \noindent Large Images.}
Image segmentation is a fundamental task in computer vision \cite{AMF+11}.
Graph-based image segmentation has been studied extensively \cite{SM00,MLM09,KNK+11}.
In these methods, each pixel is mapped into a node in a high-dimensional space (considering coordinates and intensity) that then connects to its $K$-nearest nodes.
In many applications such as in astronomy and microscopy, high-resolution images are captured with an extremely large size, up to gigapixels. Segmentation of these images usually requires pipelining, such as with deblurring as a preprocessing, so new pixels could be added for image segmentation over time. Similar to citation networks, no pixels and their edges would be deleted once they are inserted into the images.

{\bf \noindent Web Graphs.}
In a web graph with web pages as nodes and hyperlinks between pages as edges, web pages within the same community are usually densely-connected.
Clustering results on a web graph can be helpful for eliminating duplicates and recommending related pages.
There have been over 46 billion web pages on the WWW until July, 2018 \cite{Wor18}, and its size grows fast as new web pages have been constantly crawled over time.
The deletions of web pages are much less frequent and more difficult to discover than insertions.
In some cases, deleted web pages are still kept in Web graphs for analytic purposes.

All these examples require effective ways to clustering over large-scale dynamic graphs, when node/edge insertions/deletions are observed distributively and over time.
For notation convenience, we assume that we know an estimated total number of nodes in the graphs, and then node insertions and deletions are treated as insertions/deletions of its edges.
Since deletions seldom happen, we first only consider node/edge insertions, and then discuss how to include a small number of deletions in detail.
Formally, there are $s$ distributed remote sites $S_1,\cdots,S_s$ and a coordinator.
At each time point $\tau\in[1,t]$, each of these sites observes a graph update stream $\hat{E}^{\tau}_i$, defining the local graph $G^{\tau}_i(V,E^{\tau}_i=\cup_{j=1}^{\tau}\hat{E}^j_i)$ observed up to the time point $\tau$, and these sites corporate with the coordinator to generate graph clustering over the global graph $G^{\tau}(V,E^{\tau}=\cup_{i=1}^sE^{\tau}_i)$.
For simplicity, edge weights cannot be updated but an edge can be observed at different sites.
We illustrate the problem by an example in Fig. \ref{fig:illustration}.

For distributed systems, communication costs are one of the major performance measures we aim to optimize.
In this paper, we consider two well-established communication models in multi-party communication literature \cite{PVZ16}, namely the message passing and the blackboard models.
In the former model, there is a communication channel between each of the $s$ remote sites and a distinguished coordinator.
Each site can send a message to another site by first sending to the coordinator, who then forwards the message to the destination.
In the latter model, there is a broadcast channel to which a message sent is visible to all sites.
Note that both models abstract away issues of message delay, synchronization and loss and assume that each message is delivered immediately.
These assumptions can be removed by using standard techniques of timestamping, acknowledgements and re-sending, respectively.
We measure communication costs in terms of the total number of bits communicated.

Unfortunately, existing graph clustering algorithms cannot work reasonably well for the problem we considered.
In order to show the challenge, we discuss two natural methods central (\emph{CNTRL}) and static (\emph{ST}).
For every time point in $[1,t]$, \emph{CNTRL} centralizes all graph updates that are distributively arriving and then applies any centralized graph clustering algorithm. However, the total communication cost $\tilde{O}(m)$ for \emph{CNTRL} is very high, especially when the number $m$ of edges is very large.
On the other hand, for every time point in $[1,t]$, \emph{ST} applies any distributed static graph clustering algorithm on the current graph and thus adapt it to distributed dynamic setting. According to \cite{CSW+16}, the lower bounds on communication cost for distributed graph clustering in the message passing and the blackboard models are $\Omega(ns)$ and $\Omega(n+s)$, respectively, where $n$ is the number of nodes in the graph and $s$ is the number of sites. Summing over $t$ time points, the total communication cost for \emph{ST} are $\Omega(nst)$ and $\Omega(nt+st)$ resp., which could be very high especially when $t$ is very large.
Therefore, designing new algorithms for distributed dynamic graph clustering is significant and challenging because of the scarce of any valid algorithms.

\vspace{0.1in}
{\bf \noindent Contribution.}
The contribution of our work are summarized as follows.
\begin{list}{\labelitemi}{\leftmargin=0em}
\item
For the message passing model, we analyze the problem of \emph{ST} and propose an algorithm framework namely Distributed Dynamic Clustering Algorithm with Monotonicity Property (\emph{D$^2$-CAMP}), which can significantly reduce the total communication cost to $\tilde{O}(ns)$, for an $n$-node graph distributively observed at $s$ sites in a time interval $[1,t]$.
Any spectral sparsification algorithms (we will formally introduce in Sec. 2) satisfying the monotonicity property can be used in \emph{D$^2$-CAMP} to achieve the communicaiton cost.
\item
We propose an algorithm namely Distributed Dynamic Clustering Algorithm for the BLackboard model (\emph{D$^2$-CABL}) with communication cost $\tilde{O}(n+s)$ by adapting the spectral sparsification algorithm \cite{CMP16}.
\emph{D$^2$-CABL} is also a new static distributed graph clustering algorithm with nearly-optimal communication cost, the same as the iterative sampling approach \cite{LMP13} based state of the art \cite{CSW+16}.
However, it is much simpler and also works for the more complicated distributed dynamic setting.
\item
More importantly, we show that the communication costs of \emph{D$^2$-CAMP} and \emph{D$^2$-CABL} match their lower bounds $\Omega(ns)$ and $\Omega(n+s)$ up to polylogarithmic factors, respectively.
And then we prove that at every time point, \emph{D$^2$-CAMP} and \emph{D$^2$-CABL} can generate clustering results of quality nearly as good as \emph{CNTRL}.
\item
Finally, we have conducted extensive experiments on both synthetic and real-world networks to compare \emph{D$^2$-CAMP} and \emph{D$^2$-CABL} with \emph{CNTRL} and \emph{ST}, which shows that our algorithms can achieve communication cost significantly smaller than these baselines, while generating nearly the same clustering results.
\end{list}

{\bf \noindent Related Work.}
Geometric clustering has been studied by \cite{CMW07} in the distributed dynamic setting.
They presented an algorithm for k-center clustering with theoretical bounds on the clustering quality and the communication cost.
However, it is not for the graph clustering.
There have been extensive research on graph clustering in the distributed setting \cite{HYCC07,YX15,CSW+16,SZ17} where the graph is static (does not change over time) but distributed.
\cite{YX15} proposed a divide and conquer method for distributed graph clustering.
\cite{CSW+16} used spectral sparsifiers in graph clustering for two distributed communication models to reduce communication cost.
\cite{SZ17} presented a node degree based sampling scheme for distributed graph clustering, and their method does not need to compute approximate effective resistance.
However, as discussed earlier, all these methods suffer from very high communication costs, depending on the time duration, and thus cannot be used in the studied dynamic distributed clustering.
Independently, \cite{JLC18} studied distributed community detection on dynamic social networks.
However, their algorithm is not optimized for communication cost, focusing on finding overlapping clusters and only accepts unweighted graphs.
In contrast, our algorithms are optimized for communication cost. They can generate non-overlapping clusters and process both weighted and unweighted graphs.

\section{The Proposed Algorithms}

\label{sec:alg}
We first introduce spectral sparsification that we will use in subsequent algorithm design.
Recall that the message passing communication model represents distributed systems with point-to-point communication, while the blackboard model represents distributed systems with a broadcast channel, which can be used to broadcast a message to all sites.
We then propose two algorithms for different practical scenarios in Sec. \ref{sec:mp} and \ref{sec:bl}, respectively.

\vspace{0.1in}
{\bf \noindent Graph Sparsification.}
In this paper, we consider weighted undirected graphs $G(V,E,W)$ and will use $n$ and $m$ to denote the numbers of nodes and edges in $G$ respectively.
Graph sparsification is the procedure of constructing sparse subgraphs of the original graphs such that certain important property of the original graphs are well approximated.
For instance, a subgraph $H(V,E'\subseteq E)$ is called a \emph{spanner} of $G$ if for every $u,v\in V$, the shortest distance between $u$ and $v$ is at most $\alpha\geq 1$ times of their distance in $G$ \cite{PS89}.
Let $A_G$ be the adjacency matrix of $G$.
That is, $(A_G)_{u,v}=W(u,v)$ if $(u,v)\in E$ and zero otherwise.
Let $D_G$ be the degree matrix of $G$ defined as $(D_G)_{u,v}=\sum_{v\in V}W(u,v)$, and zero otherwise.
Then the unnormalized \emph{Laplacian matrix} and \emph{normalized Laplacian matrix} of $G$ are defined as $L_G=D_G-A_G$ and $\mathcal{L}_G=D_G^{-1/2}L_GD_G^{-1/2}$, resp..
\cite{ST11} introduced \emph{spectral sparsification}:
a $(1+\epsilon)$-\emph{spectral sparsifier} for $G$ is a subgraph $H$ of $G$, such that for every $x\in R^n$, the inequality
$(1-\epsilon)x^TL_Gx\leq x^TL_Hx\leq (1+\epsilon)x^TL_Gx$ holds.
There is a rich literature on improving the trade-off between the size of spectral sparsifiers and the construction time, e.g. \cite{SS11,ZLO15,LS17}.
Recently, \cite{LS17} proposed the state-of-the-art algorithm to construct a $(1+\epsilon)$-spectral sparsifier of optimal size $O(n/\epsilon^2)$ (up to a constant factor) in nearly linear time $\tilde{O}(m)$.

\subsection{The Message Passing Model}
\label{sec:mp}
Because spectral sparsifiers have much fewer edges than the original graphs but can preserve cut-based clustering and spectrum information of the original graphs \cite{SS11}, we propose an algorithm framework as follows.
At each time point $\tau$, each site $S_i$ first constructs a spectral sparsifier $H_i^{\tau}$ for the local graph $G_i^{\tau}(V,E_i^{\tau})$, and then transmits the much smaller $H_i^{\tau}$, instead of $G_i^{\tau}$ itself, to the coordinator.
Upon receiving the spectral sparsifier $H_i^{\tau}$ from every site at the time $\tau$, the coordinator first takes their union $H^{\tau}=\cup_{i=1}^sH_i^{\tau}$ and then applies a standard centralized graph clustering algorithm, e.g., the spectral clustering algorithm \cite{NJW01}, on $H^{\tau}$ to get the clustering $C^{\tau}$.
This process is repeated at the next time point $\tau+1$ to get the clustering $C^{\tau+1}$ until $t$.

However, simply re-constructing spectral sparsifiers from scratch at every time point does not provide any bound on the size of the updates to the previous spectral sparsifiers $H^{\tau-1}_i$ for obtaining $H^{\tau}_i$ at every time point $\tau$, and thus needs to communicate the entire spectral sparsifiers $H^{\tau}_i$ of size $O(n)$ at every time point $\tau$.
Summing over all $s$ sites and all $t$ time points, the total communication cost is $\tilde{O}(nst)$.

It is natural to consider algorithms for dynamically maintaining spectral sparsifiers in dynamic computational models \cite{ADK+16,KL13,KLM+14}.
Unfortunately, applying them also does not provide such a bound, incurring the same communication cost!
To see this, the key of (algorithms in) dynamic computational models is a data structure for dynamically maintaining the result of a computation while the underlying input data is updated periodically.
For instance, dynamic algorithms \cite{ADK+16}, after each update to the input data, are allowed to process the update to compute the new result within a fast time; online algorithms \cite{KL13} allow to process the input data that are revealed step by step; and streaming algorithms \cite{KLM+14} impose a space constraint while processing the input data that are revealed step by step.
The main principle of all these computational models is on efficiently processing the dynamically changing input data, instead of bounding the size of the updates to the previous output result over time.

We define a new type of spectral sparsification algorithms, which can provide such a bound, and is defined as follows.
\begin{definition}
For an $n$-node graph $G(V,E$=$\{e_1,\cdots,e_m\})$, let $G(V,E_i$=$\{e_1,\cdots,e_i\})$ be the graph consisting of the first $i$ edges.
A spectral sparsification algorithm is called a Spectral Sparsification Algorithm with Monotonicity Property (\emph{S$^2$AMP}), if the spectral sparsifers $H_1,\cdots,H_m$, constructed for $G_1,\cdots,G_m$, respectively, satisfy that (1) $H_1\subseteq\cdots \subseteq H_m$; and (2) $H_m$ has size $\tilde{O}(n)$.
\end{definition}

We show that, by using any \emph{S$^2$AMP} in the algorithm framework mentioned above, we can reduce the total communication cost from $\tilde{O}(nst)$ to $\tilde{O}(ns)$, removing a factor of $t$.
We refer to the resultant algorithm framework as Distributed Dynamic Clustering Algorithm with Monotonicity Property (\emph{D$^2$-CAMP}).
The intuition for the significant reduction in the total communication cost is that, the monotonicity property guarantees that, for every time point $\tau\in [1,t]$, the constructed spectral sparsifiers $H^{\tau}_i$ is a superset of $H^{\tau-1}_i$ at the previous time point $\tau-1$.
Then, we only need to transmit edges in $H^{\tau}_i$ and at the same time not in $H^{\tau-1}_i$ to the coordinator for maintaining $H^{\tau}_i$.
Every communicated bit transmitted at the time point $t$ is used at all subsequent time points $\{\tau+1,\cdots,t\}$, and thus no communication is ``wasted''.
Furthermore, we show that by only switching an arbitrary spectral sparsification algorithm to \emph{S$^2$AMP}, the total communication cost $\tilde{O}(ns)$ achieved has been optimal, up to a polylogarithmic factor.
That is, we cannot design another algorithm with communication cost smaller than \emph{D$^2$-CAMP} by a polylogarithmic factor.

We summarize the results in Theorem \ref{thm:mp}.
For every node set $S\subseteq V$ in $G$, let its \emph{volume} and \emph{conductance} be $vol_G(S)=\sum_{u\in S,v\in V}W(u,v)$ and $\phi_G(S)=(\sum_{u\in S,v\in V-S}W(u,v))/vol_G(S)$, respectively.
Intuitively, a small value of conductance $\phi(S)$ implies that nodes in $S$ are likely to form a cluster.
A collection of subsets $A_1,\cdots,A_k$ of nodes is called a \emph{(k-way) partition} of $G$ if (1) $A_i\cap A_j=\emptyset$ for $1\leq i\not=j\leq k$; and (2) $\cup_{i=1}^kA_i=V$.
The \emph{k-way expansion constant} is defined as $\rho(k)=\min_{partition A_1,\cdots,A_k}\max_{i\in[1,k]}\phi(A_i)$.
The eigenvalues of $\mathcal{L}_G$ are denoted as $\lambda_1(\mathcal{L}_G)\leq \cdots\leq \lambda_n(\mathcal{L}_G)$.
The high-order Cheeger inequality shows that $\lambda_k/2\leq \rho(k)\leq O(k^2)\sqrt{\lambda_k}$ \cite{LGT14}.
A lower bound on $\Upsilon_G(k)=\lambda_{k+1}/\rho(k)$ implies that, $G$ has exactly $k$ well-defined clusters \cite{PSZ15}.
It is because a large gap between $\lambda_{k+1}$ and $\rho(k)$ guarantees the existence of a k-way partition $A_1,\cdots,A_k$ with bounded $\phi(A_i)\leq \rho(k)$, and that any $(k+1)$-way partition $A_1,\cdots,A_{k+1}$ contains a subset $A_i$  with significantly higher conductance $\rho(k+1)\geq \lambda_{k+1}/2$ compared with $\rho(k)$.
For any two sets $X$ and $Y$, the symmetric difference of $X$ and $Y$ is defined as $X\Delta Y=(X-Y)\cup(Y-X)$.
To prove Theorem \ref{thm:mp}, we will use the following lemma and theorems.

\begin{lemma}{\cite{CSW+16}}
\label{thm:conductance}
Let $H$ be a $(1+\epsilon)$-spectral sparsifier of $G(V,E)$ for some $\epsilon\leq 1/3$.
For all node sets $S\subseteq V$, the inequality $0.5\cdot\phi_G(S)\leq \phi_H(S)\leq 2\cdot\phi_G(S)$ holds.
\end{lemma}

\begin{theorem}{\cite{CSW+16}}
\label{thm:clb}
Let $G$ be an $n$-node graph and the edges of $G$ are distributed amongst $s$ sites.
Any algorithm that correctly outputs a constant fraction of each cluster in $G$ requires $\Omega(ns)$ bits of communications.
\end{theorem}

\begin{theorem}
\label{thm:spectralclustering}
\cite{PSZ15} Given a graph $G$ with $\Upsilon_G(k)=\Omega(k^3)$ and an optimal partition $S_1,\cdots,S_k$ achieving $\rho(k)$ for some positive integer $k$, the spectral clustering algorithm can output partition $A_1,\cdots,A_k$ such that, for every $i\in [1,k]$, the inequality $vol(A_i\Delta S_i)=O(k^3\Upsilon^{-1}vol(S_i))$ holds.
\end{theorem}

\begin{theorem}[The Message Passing model]
\label{thm:mp}
For every time point $\tau\in [1,t]$, suppose that $G^{\tau}$ satisfies that $\Upsilon(k)=\Omega(k^3)$ and there is an optimal partition $P_1,\cdots,P_k$ which achieves $\rho(k)$ for some positive integer $k$, D$^2$-CAMP can output partition $A_1,\cdots,A_k$ at the coordinator such that for every $i\in [1,k]$, $vol(A_i\Delta P_i)=O(k^3\Upsilon^{-1}vol(P_i))$ holds.
Summing over all $t$ time points, the total communication cost is $\tilde{O}(ns)$. It is optimal up to a polylogarithmic factor.
\end{theorem}

\proof
We start by proving that for every time point $\tau\in[1,t]$, the structure $H^{\tau}$ constructed at the coordinator is a $(1+\epsilon)$-spectral sparsifier of the graph $G^{\tau}$ received up to the time point $t$.
By the monotonicity property of a \emph{S$^2$AMP}, for every $i\in [1,s]$, $H^{\tau}_i$ is a $(1+\epsilon)$-spectral sparsifier of the graph $G^{\tau}_i(V,E^{\tau}_i)$.
The decomposability of spectral sparsifiers states that the union of spectral sparisifiers of some graphs is a spectral sparsifier for the union of the graphs \cite{SZ17}.
Then by this property, the union of $H^{\tau}=\cup_{i=1}^sH^{\tau}_i$ obtained at the coordinator is a $(1+\epsilon)$-spectral sparsifier of the graph $G^{\tau}=\cup_{i=1}^sG^{\tau}_i$.

Now we prove that for every time point $\tau\in[1,t]$, if $G^{\tau}$ satisfies that $\Upsilon_{G^{\tau}}(k)=\Omega(k^3)$, $H^{\tau}$ also satisfies that $\Upsilon_{H^{\tau}}(k)=\Omega(k^3)$.
By the definition of $\Upsilon$, it suffices to prove that $\rho_{H^{\tau}}(k)=\Theta(\rho_{H^{\tau}}(k))$ and $\lambda_{k+1}(\mathcal{L}_{H^{\tau}})=\Theta(\lambda_{k+1}(\mathcal{L}_{G^{\tau}}))$.
The former follows from that for every $i\in [1,k]$, the inequality
\begin{equation*}
0.5\cdot\phi_{G^{\tau}}(S_i)\leq \phi_{H^{\tau}}(S_i)\leq 2\cdot\phi_{G^{\tau}}(S_i)
\end{equation*}
holds, according to Lemma \ref{thm:conductance}.
According to the definition of $(1+\epsilon)$-spectral sparsifier and simple math, it holds for every vector $x\in R^n$ that
\begin{equation*}
\begin{aligned}
(1-\epsilon)x^TD_{G^{\tau}}^{-1/2}L_{G^{\tau}}D_{G^{\tau}}^{-1/2}x\leq x^TD_{G^{\tau}}^{-1/2}L_{H^{\tau}}D_{G^{\tau}}^{-1/2}x\\\leq (1+\epsilon)x^TD_{G^{\tau}}^{-1/2}L_{G^{\tau}}D_{G^{\tau}}^{-1/2}x.
\end{aligned}
\end{equation*}
By the definition of normalized graph Laplacian $\mathcal{L}_G$, and the fact that for every vector $y\in R^n$,
\begin{equation*}
0.5\cdot y^TD_{G^{\tau}}^{-1}y\leq y^TD_{H^{\tau}}^{-1}y\leq 2y^TD_{G^{\tau}}^{-1}y,
\end{equation*}
we have that for every $i\in [1,n]$,
\begin{equation*}
\lambda_i(\mathcal{L}_{H^{\tau}})=\Theta(\lambda_i(\mathcal{L}_{G^{\tau}})),
\end{equation*}
which implies that $\lambda_{k+1}(\mathcal{L}_{H^{\tau}})=\Theta(\lambda_{k+1}(\mathcal{L}_{G^{\tau}}))$.
Then we can apply the spectral clustering algorithm on $H^{\tau}$ to get the desirable properties, according to Theorem \ref{thm:spectralclustering}.

For the upper bound on the communication cost, by the monotonicity property of a \emph{S$^2$AMP}, each site only needs to transmit $\tilde{O}(n)$ number of edges over all $t$ time points.
Summing over all $s$ sites, the total communication cost is $\tilde{O}(ns)$.

For the lower bound, we show the following statement.
For every time point $\tau\in [1,t]$, suppose $G^{\tau}$ satisfies that $\Upsilon(k)=\Omega(k^3)$ and there is an optimal partition $P_1,\cdots,P_k$ which achieves $\rho(k)$ for positive integer $k$, in the message passing model there is an algorithm which can output $A_1,\cdots,A_k$ at the coordinator, such that for every $i\in [1,k]$, $vol(A_i\Delta P_i)=\Theta(vol(P_i))$ holds.
Then the algorithm requires $\Omega(ns)$ total communication cost over $t$ time points.

Consider any time point $\tau$.
We assume by contradiction that there exists an algorithm which can output $A_1,\cdots,A_k$ in $G^{\tau}$ at the coordinator, such that for every $i\in [1,k]$, $vol(A_i\Delta P_i)=\Theta(vol(P_i))$ holds, using $o(ns)$ bits of communications.
Then the algorithm can be used to solve a corresponding graph clustering problem in the distributed but static setting using $o(ns)$ bits of communications.
This contradicts Theorem \ref{thm:clb}, and then completes the proof.
\qed
\vspace{0.1in}

Combining Theorems \ref{thm:spectralclustering} and \ref{thm:mp}, \emph{D$^2$-CAMP} could generate clustering of quality asymptotically the same as \emph{CNTRL}.
We stress that the monotonicity property in general can be helpful for improving the communication efficiency over distributed dynamic graphs.
In Sec. \ref{sec:discussion}, we will discuss a new application which also benefits from the property.

As mentioned earlier, any \emph{S$^2$AMP} algorithm can be plugged in \emph{D$^2$-CAMP}, e.g., the online sampling technique \cite{CMP16}.
But the resultant algorithm becomes a randomized algorithm which succeeds w.h.p. because the constructed subgraphs are spectral sparsifiers w.h.p.
Another \emph{S$^2$AMP} algorithm is the online-BSS algorithm \cite{BSS12,CMP16}, which has a slightly smaller communication cost (by a logarithmic factor) but requires larger memory and is more complicated.

\subsection{The Blackboard Model}
\label{sec:bl}

How to efficiently exploit the broadcast channel in the blackboard model to reduce the communication complexity in distributed graph clustering is non-trivial.
For example, \cite{CSW+16} proposed to construct $O(\log n)$ spectral sparisifers as a chain in the blackboard based on the iterative sampling technique \cite{LMP13}.
Each spectral sparsifier in the chain is a spectral sparsifer of its following sparsifier.
However, the technique fails to extend to the dynamic setting, as each graph update could incur a large number of updates in the maintained spectral sparsifiers, especially for those in the latter part of the chain.

We propose a simple algorithm called Distributed Dynamic Clustering Algorithm for the BLackboard model (\emph{D$^2$-CABL}), based on adapting Cohen et al.'s algorithm \cite{CMP16}.
The basic idea is that every site corporates with each other to construct a spectral sparsifier $H^{\tau}$ for $G^{\tau}(V,E^{\tau})$ at each time point $\tau$ in the blackboard.

\vspace{-0.05in}
\begin{algorithm}[h]
\caption{\emph{D$^2$-CABL} at Time Point $\tau$}
\KwIn{The incidence matrix $B^{\tau-1}$, new edges $\hat{E}^{\tau}$ coming at $\tau$, $\delta>0$, $\epsilon\in (0,1/3)$}
\KwOut{The incidence matrix $B^{\tau}$}
$\lambda\leftarrow \delta/\epsilon$;
$c\leftarrow 8\log n/\epsilon^2$\;
$B'\leftarrow B^{\tau-1}$\;
\For{$e\in \hat{E}^{\tau}$} {
    $l=(1+\epsilon)b(e)^T(B'^TB'+(\delta/\epsilon) I)^{-1}b(e)$\;
    $p\leftarrow\min\{cl,1\}$\;
    $B'\leftarrow[B';b(e)/\sqrt{p}]$ with probability $p$\;
}
\Return{$B^{\tau}\leftarrow B'$}\;
\label{alg:bl}
\end{algorithm}
\vspace{-0.1in}

The edge-node incidence matrix $B_{m\times n}$ of $G$ is defined as $B(e,v)=1$ if $v$ is $e$'s head, $B(e,v)=-1$ if $v$ is $e$'s tail, and zero otherwise.
At the beginning, the parameters $\delta$ and $\epsilon$ of the algorithm are set by a distinguished site and then sent to every site, and the blackboard has an empty spectral sparsifier $H^0$, or equivalently an empty incidence matrix $B^0$ of dimension $0\times n$.
Consider the time point $\tau$.
Suppose that at the previous time point $\tau-1$, the incidence matrix $B^{\tau-1}$ for $H^{\tau-1}$ was in the blackboard.
For each newly observed edge $e\in \hat{E}^{\tau}$ at the time point $\tau$, the site $S_i$ observing $e$ computes the online ridge leverage score $l=(1+\epsilon)b(e)^T(B'^TB'+(\delta/\epsilon) I)^{-1}b(e)$ by accessing the incidence matrix $B'$ currently in the blackboard, where $b(e)$ is an $n$-dimensional vector with all zeroes except that the entries corresponding to $e$'s head and tail are 1 and -1, resp..

Let the sampling probability $p=\min\{(8\log n/\epsilon^2)l,1\}$.
With probability $p$, $e$ is sampled, or discarded otherwise.
If $e$ is sampled, the site $S_i$ transmits the rescaled vector $b(e)/\sqrt{p}$ corresponding to $e$ to the blackboard to append it at the end of $B'$.
After all the newly observed edges $\hat{E}^{\tau}$ at the time point $\tau$ at all the sites are processed, $B^{\tau}$ for $H^{\tau}$ will be in the blackboard.
Then the coordinator applies any standard graph clustering algorithm, e.g. \cite{NJW01}, on $H^{\tau}$ to get the clustering $C^{\tau}$.
The process is repeated for every subsequent time point until $t$.
The algorithm is summarized in Alg. 1.

Our results for the blackboard model are summarized in Theorem \ref{thm:bl}.
To prove Theorem \ref{thm:bl}, first it follows from \cite{CMP16} that the constructed subgraph in the blackboard for every time point $\tau$ is a spectral sparsifier for the graph $G^{\tau}$ w.h.p..
Then the rest of the proof is the same as the proof of Theorem \ref{thm:mp}.
In the algorithm, processing an edge requires only $B'$, which is in the blackboard and visible to every site.
Therefore, each site can process its edges locally and only transmit the sampled edges to the blackboard.
The total communication cost is $\tilde{O}(n+s)$, because the size of the constructed spectral sparsifier is $\tilde{O}(n)$ and each site has to transmit at least one bit of information.
It is easy to see this communication cost is optimal up to polylogarithmic factors, because even only for one time point, the clustering result itself has $\Omega(n)$ bits of information and each site has to transmit at least one bit of information.

\begin{theorem}[The Blackboard model]
\label{thm:bl}
For every time point $\tau\in [1,t]$, suppose that $G^{\tau}$ satisfies that $\Upsilon(k)=\Omega(k^3)$ and there is an optimal partition $P_1,\cdots,P_k$ which achieves $\rho(k)$ for some positive integer $k$, w.h.p. \emph{D$^2$-CABL} can output partition $A_1,\cdots,A_k$ at the coordinator such that for every $i\in [1,k]$, $vol(A_i\Delta P_i)=O(k^3\Upsilon^{-1}vol(P_i))$ holds.
Summing over $t$ time points, the total communication cost is $\tilde{O}(n+s)$. It is optimal up to a polylogarithmic factor.
\end{theorem}


\emph{D$^2$-CABL} can also work in the distributed static setting by considering that there is only one time point, at which all graph information comes together.
As mentioned earlier, it is a brand new algorithm with nearly-optimal communication complexity, the same as the state-of-the-art algorithm \cite{CSW+16}.
But our algorithm is much simpler without having to maintain a chain of spectral sparsifiers.
Another advantage is the simplicity that one algorithm works for both distributed settings.
The computational complexity for computing the online ridge leverage score for each edge in Alg. 1 is $O(n^2m)$.
To save computational cost, we can batch process in every site new edges $\hat{E}^{\tau}_i$ observed at each time point $\tau$ in a batch of $O(n)$.
By using the Johnson-Linderstrauss random projection trick \cite{SS11}, we can approximate online ridge leverage scores for a batch of $O(n)$ edges in $\tilde{O}(n\log m)=\tilde{O}(n)$ time, and then sample all edges together according to the computed scores.

\section{Discussions}
\label{sec:discussion}

{\bf Another Application of the Monotonicity Property.}
Consider the same computational and communication models.
When the queries posed at the coordinator are changed to approximate shortest path distance queries between two given nodes, we use graph spanners \cite{PS89,ADD+93} to sparsify the original graphs while well approximating all-pair shortest path distances in the original graphs.

We now describe the algorithm.
In the message passing model, at each time point $t$ each site $S_i$ first constructs a graph spanner $Q^{\tau}_i$ of the local graph $G^{\tau}_i(V,E^{\tau}_i)$ using a \emph{D$^2$-CAMP} for constructing graph spanners \cite{Elk11}, and then transmits $Q^{\tau}_i$ to the coordinator.
Upon receiving $Q^{\tau}_i$ from every site, the coordinator first takes their union $Q^{\tau}=\cup_{i=1}^sQ^{\tau}_i$ and then applies a point-to-point shortest path algorithm (e.g., Dijkstra's algorithm \cite{Dij59}) on $Q^{\tau}$ to get the shortest distance between the two nodes at the time point $\tau$.
This process is repeated for every $\tau\in [1,t]$.
The theoretical guarantees of the algorithm are summarized in Theorem \ref{thm:spanner}, and its proof is in Sec. 3 of Appendix.

\begin{theorem}
\label{thm:spanner}
Given two nodes $u,v\in V$ and an integer $k>1$, for every time point $\tau\in [1,t]$, the proposed algorithm can answer approximate shortest distance between $u$ and $v$ in $G^{\tau}$ no larger than $2k-1$ times of their actual shortest distance at the coordinator in the message passing model.
Summing over $t$ time points, the total communication cost is $\tilde{O}(n^{1+1/k}s)$.
\end{theorem}

{\bf \noindent Dynamic Graph Streams.}
When the graph update stream observed at each site is a fully dynamic stream containing a small number of node/edge deletions, we present a simple trick which enables that our algorithms still have good performance.
We observe that the spectral sparsifiers can probably keep unchanged, when there is only a small number of deletions.
This is reasonable because spectral sparsifiers are sparse subgraphs which could contain much smaller edges than the original graphs.
When the number of deletions is small, the deletions may not affect the spectral sparsifiers at all.
Even when the deletions lead to small changes in the spectral sparsifiers, there is a high probability that the clustering is not changed significantly.
Therefore, in order to save communication and computation, we can ignore and do not process or transmit these deletions while still approximately preserving the clustering.
We experimentally confirm the effects of this thick in the experiment section.

\section{Experiments}
\label{sec:experiment}
In this section, we present the experimental results that we conducted on both synthetic and real-life datasets, where we compared the proposed algorithms \emph{D$^2$-CAMP} and \emph{D$^2$-CABL} with baseline algorithms \emph{CNTRL} and \emph{ST}.
For \emph{ST}, we used the distributed static graph clustering algorithms \cite{CSW+16} in the message passing and the blackboard models, and refer the resultant algorithms as \emph{STMP} and \emph{STBL}, respectively.
For measuring the quality of the clustering results, we used the normalized cut value (NCut) of the clustering \cite{SZ17}.
A smaller value of NCut implies a better clustering while a larger value of NCut implies a worse clustering.
For simplicity, we used the total number of edges communicated as the communication cost, which approximates the total number of bits by a logarithmic factor.
We implemented all five algorithms in Matlab programs, and conducted the experiments on a machine equipped with Intel i7 7700 2.8GHz CPU, 8G RAM and 1T disk storage.


The details of the datasets we used in the experiments are described as follows.
The \emph{Gaussians} dataset consists of 800 nodes and 47,897 edges.
Each point from each of four clusters is sampled from an isotropic Gaussians of variance 0.01.
We consider each point to be a node in constructing the similarity graph.
For every two nodes $u$ and $v$ such that one is among the 100-nearest points of the other, we add an edge of weight $W(u,v)=exp\{-||u-v||_2^2/2\sigma^2\}$ with $\sigma=1$.
The number $k$ of clusters is 4.
For the \emph{Sculpture} dataset, we used a $22\times 30$ version of a photo of The Greek Slave
\footnote{http://artgallery.yale.edu/collections/objects/14794},
and it contains 1980 nodes and 61,452 edges.
We consider each pixel to be a node by mapping each pixel to a point in $R^5$, i.e. $(x,y,r,g,b)$, where the last three coordinates are the RGB values.
For every two nodes $u$ and $v$ such that $u$ ($v$) is among the 80-nearest points of $v$ ($u$), we add an edge of weight $W(u,v)=exp\{-||u-v||_2^2/2\sigma^2\}$ with $\sigma=20$.
The number $k$ of clusters is 3.

In the problem studied, the site and the time point each edge comes is arbitrary.
Therefore, we make that the edges of nodes with smaller $x$ coordinates have smaller arrival times than the edges of nodes with larger $x$ coordinates.
Intuitively, this results in that the edges of nodes on the left side come before the edges of nodes on the right side.
This helps us to easily monitor the changing of the clustering results.
Independently, the site every edge comes is randomly picked from the interval $[1,s]$.


\begin{figure*}[ht]
\center
\begin{subfigure}[ht!]{.5\columnwidth}
     \centerline{\includegraphics[width=\columnwidth]{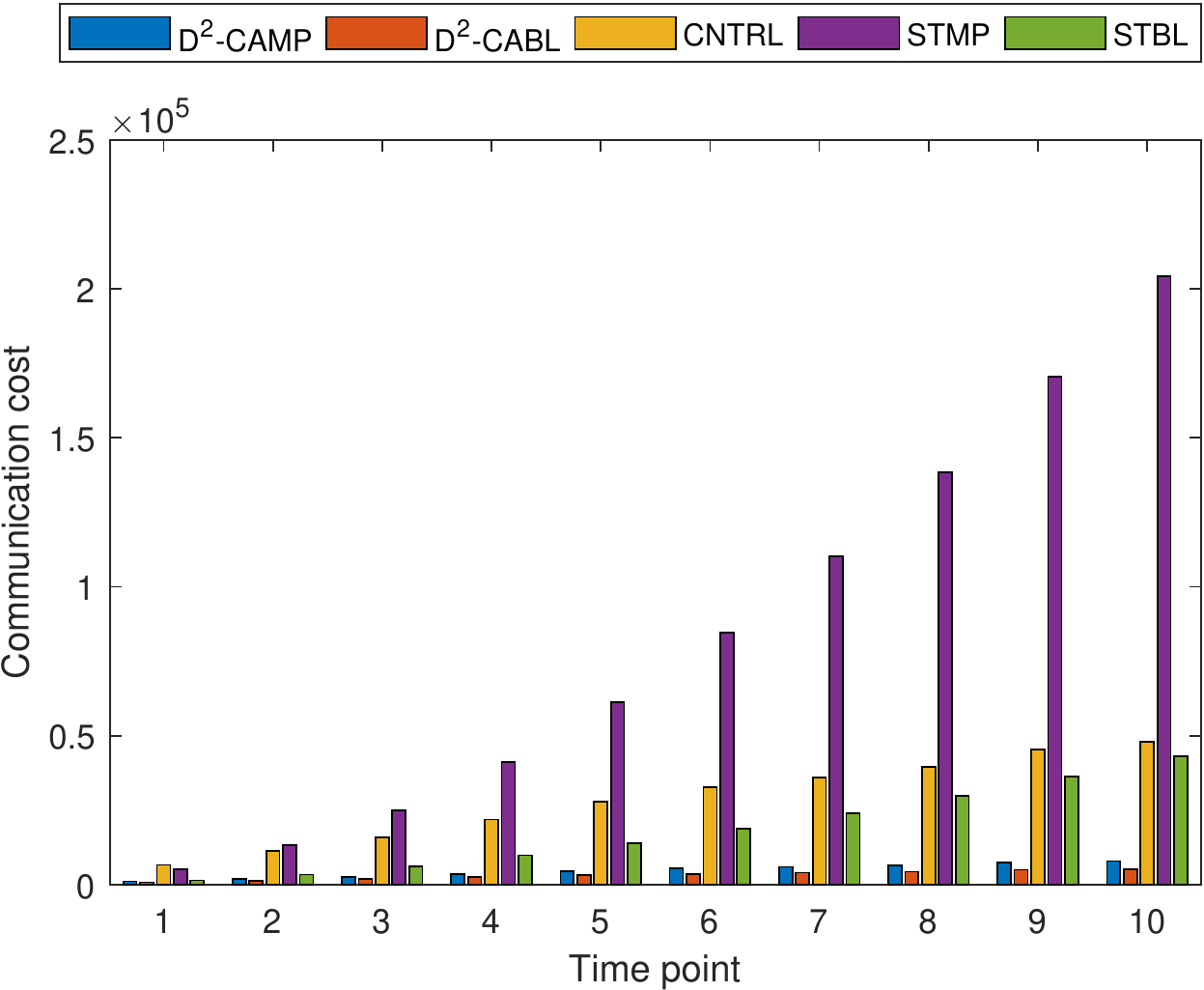}}
     \caption{\small Comm. on Gaussians dataset}
     \label{fig:gcc}
\end{subfigure}
\begin{subfigure}[ht!]{.5\columnwidth}
     \centerline{\includegraphics[width=\columnwidth]{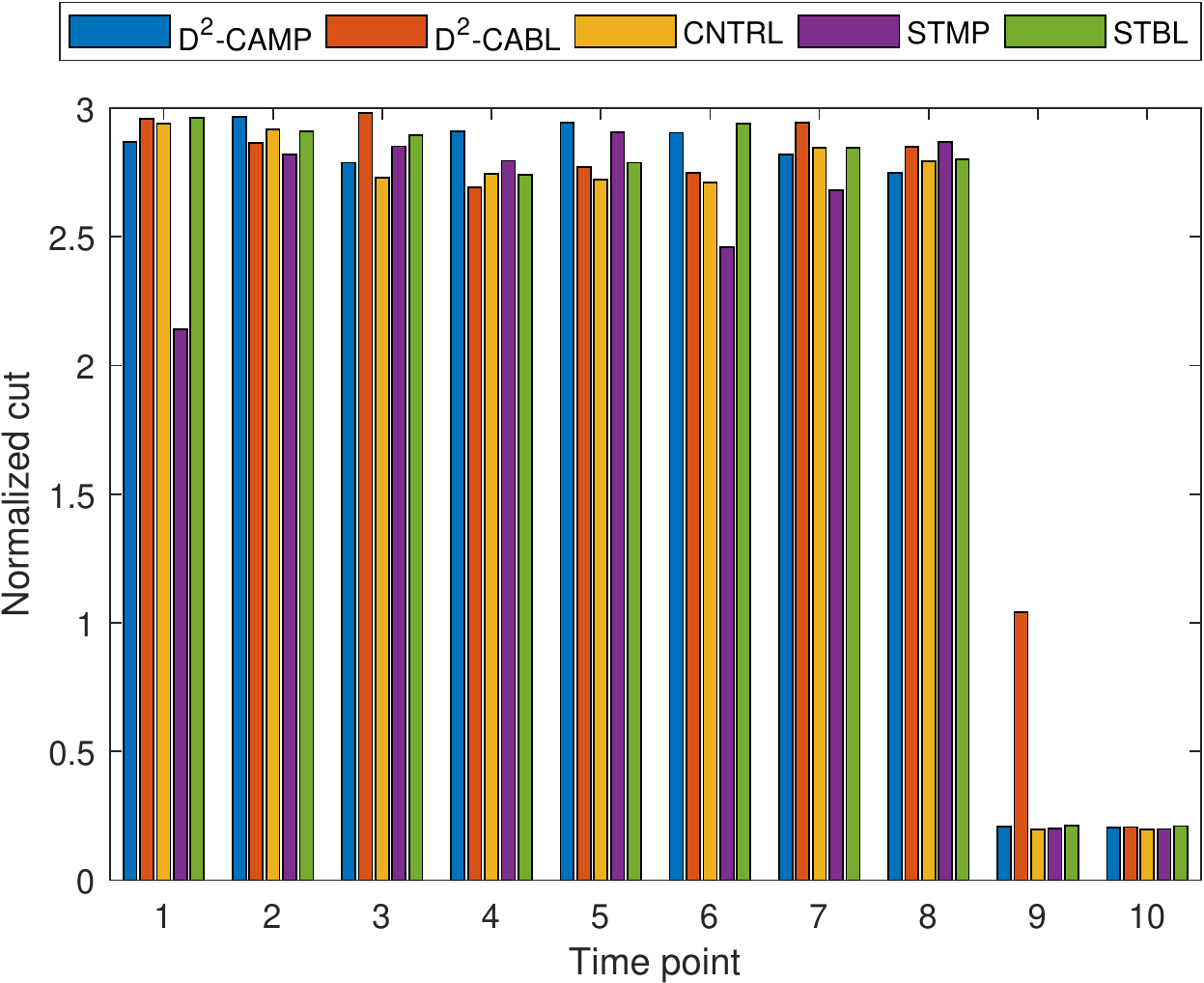}}
     \caption{\small NCut on Gaussians dataset}
     \label{fig:gnc}
\end{subfigure}
\begin{subfigure}[ht!]{.5\columnwidth}
    \includegraphics[width=\columnwidth]{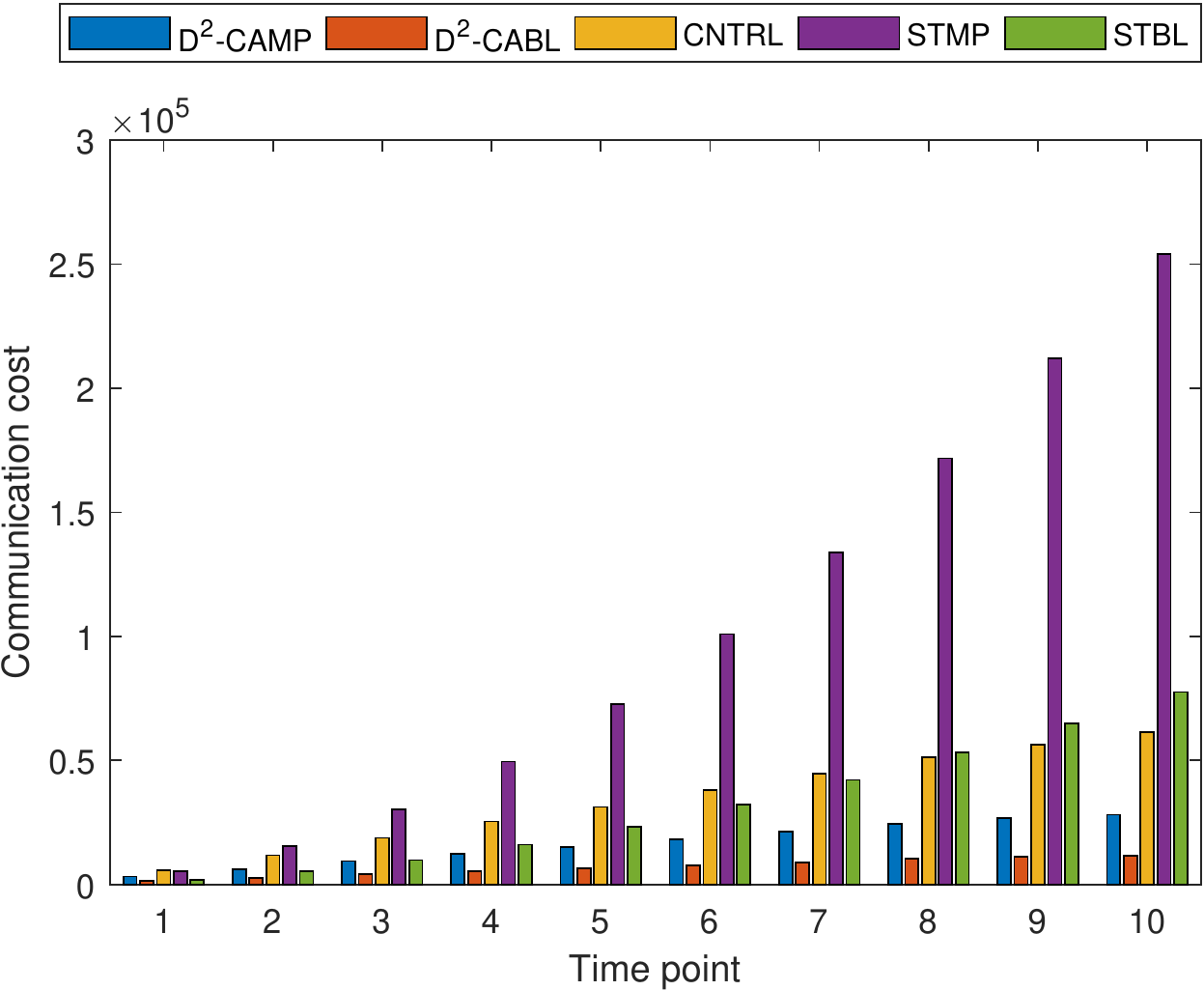}
    \caption{\small Comm. on Sculpture dataset}
    \label{fig:scc}
\end{subfigure}
\begin{subfigure}[ht!]{.5\columnwidth}
    \includegraphics[width=\columnwidth]{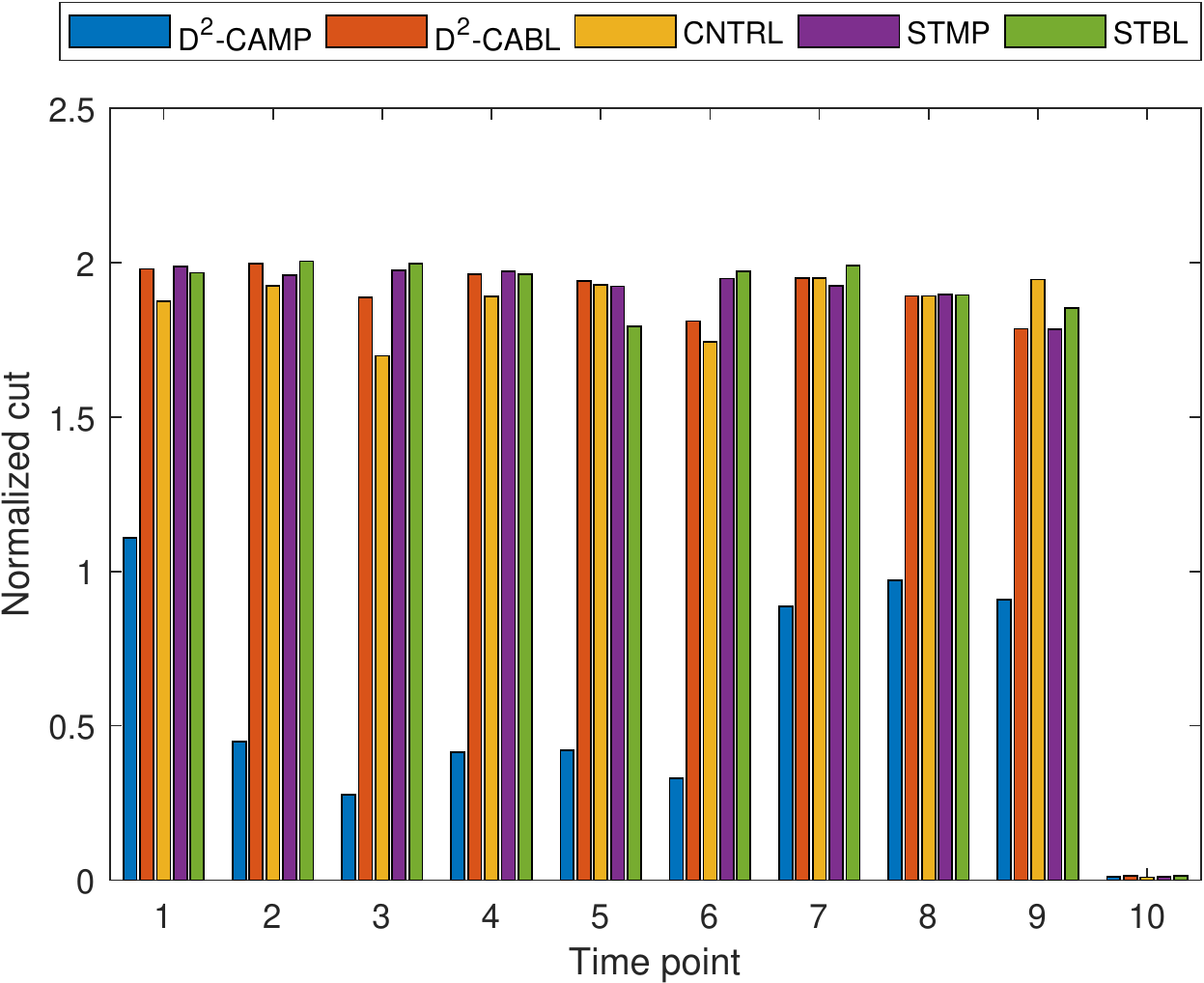}
    \caption{\small NCut on Sculpture dataset}
    \label{fig:snc}
\end{subfigure}


\begin{subfigure}[ht!]{.5\columnwidth}
    \includegraphics[width=\columnwidth]{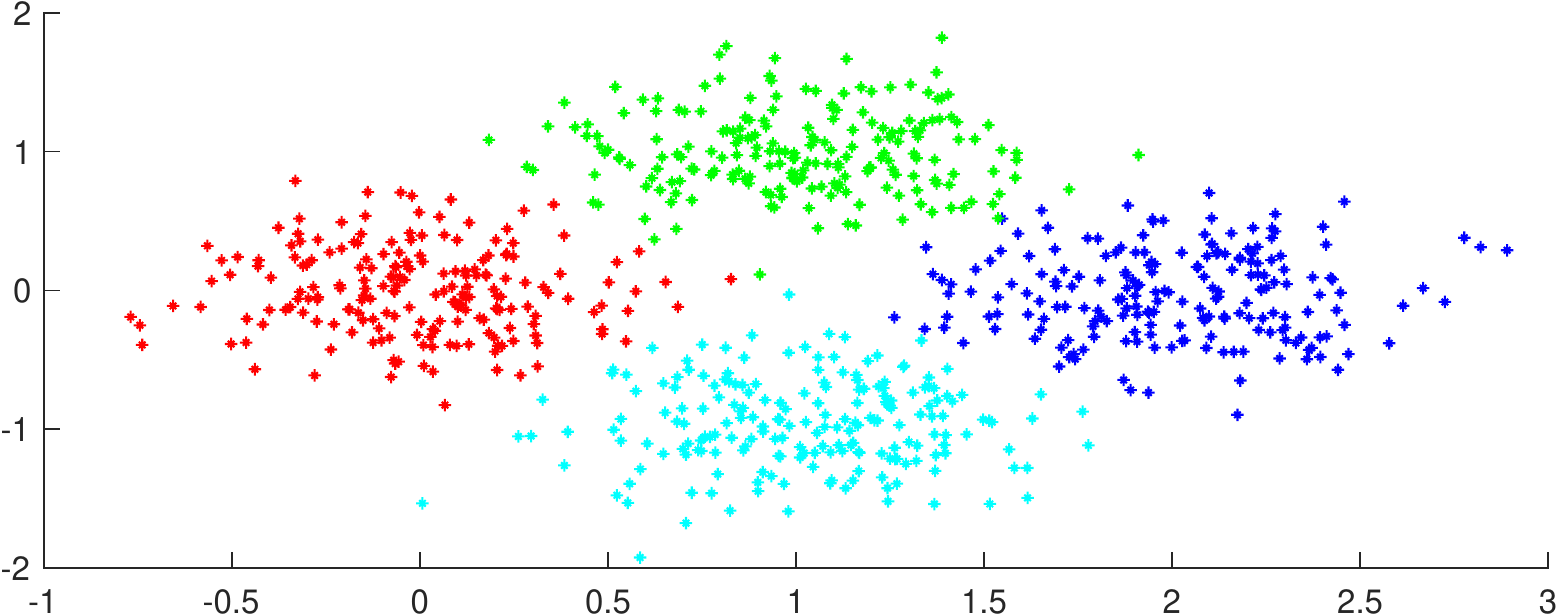}
    \caption{\small \emph{CNTRL} on Gaussians dataset at time point 9}
    \label{fig:baseline1_9_g}
\end{subfigure}
\begin{subfigure}[ht!]{.5\columnwidth}
    \includegraphics[width=\columnwidth]{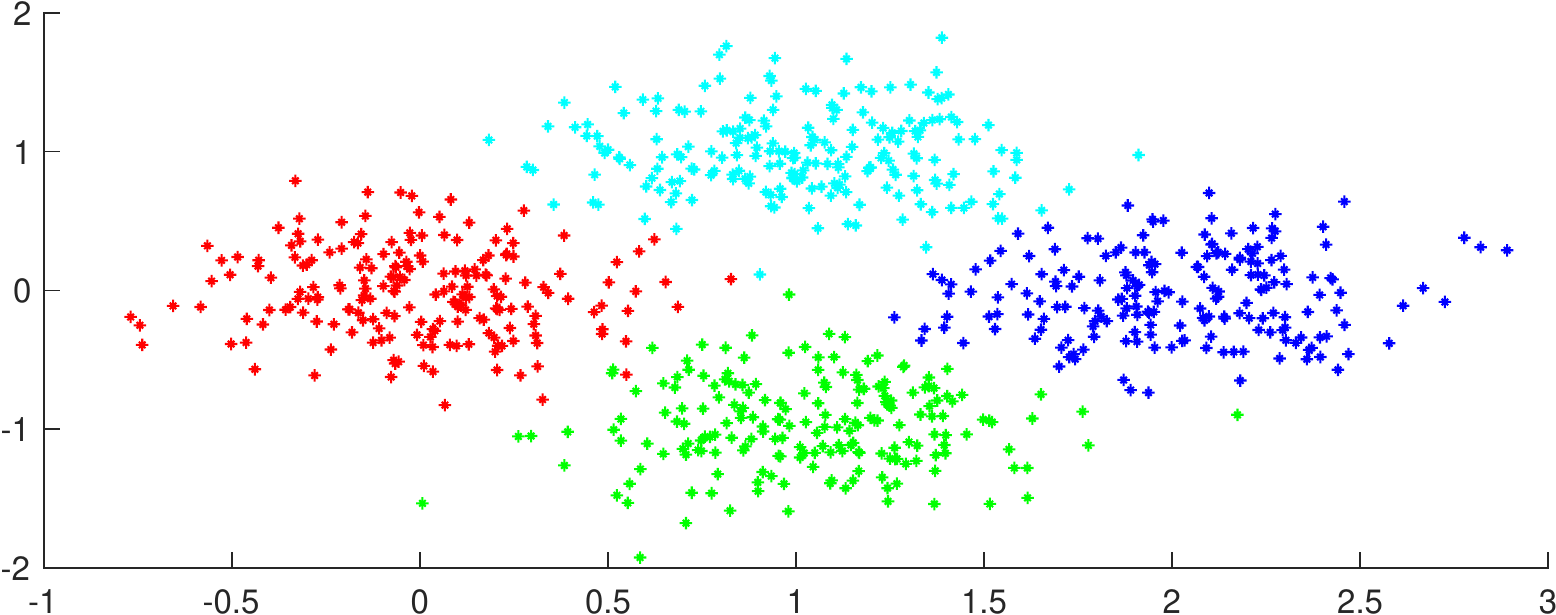}
    \caption{\small \emph{D$^2$-CAMP} on Gaussians dataset at time point 9}
    \label{fig:specalgo_s30_9_g}
\end{subfigure}
\begin{subfigure}[ht!]{.5\columnwidth}
    \includegraphics[width=\columnwidth]{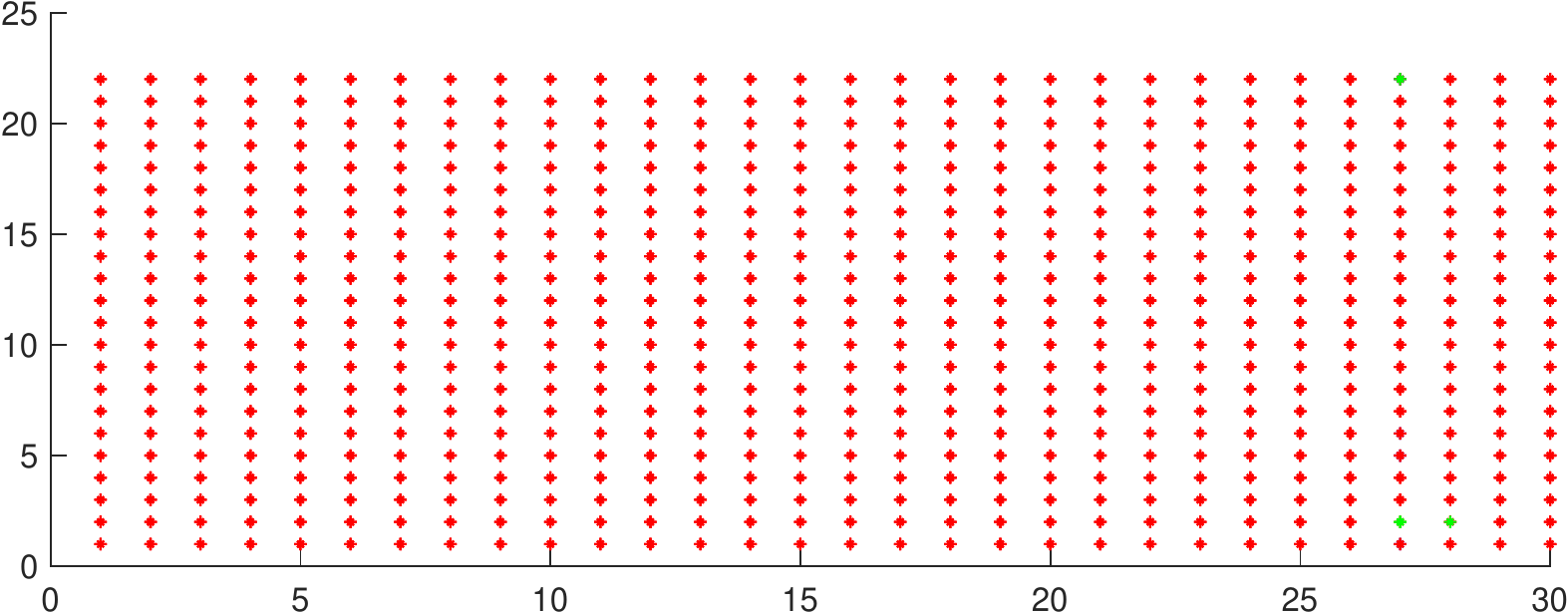}
    \caption{\small \emph{CNTRL} on Sculpture dataset at time point 9}
    \label{fig:baseline1_9_s}
\end{subfigure}
\begin{subfigure}[ht!]{.5\columnwidth}
    \includegraphics[width=\columnwidth]{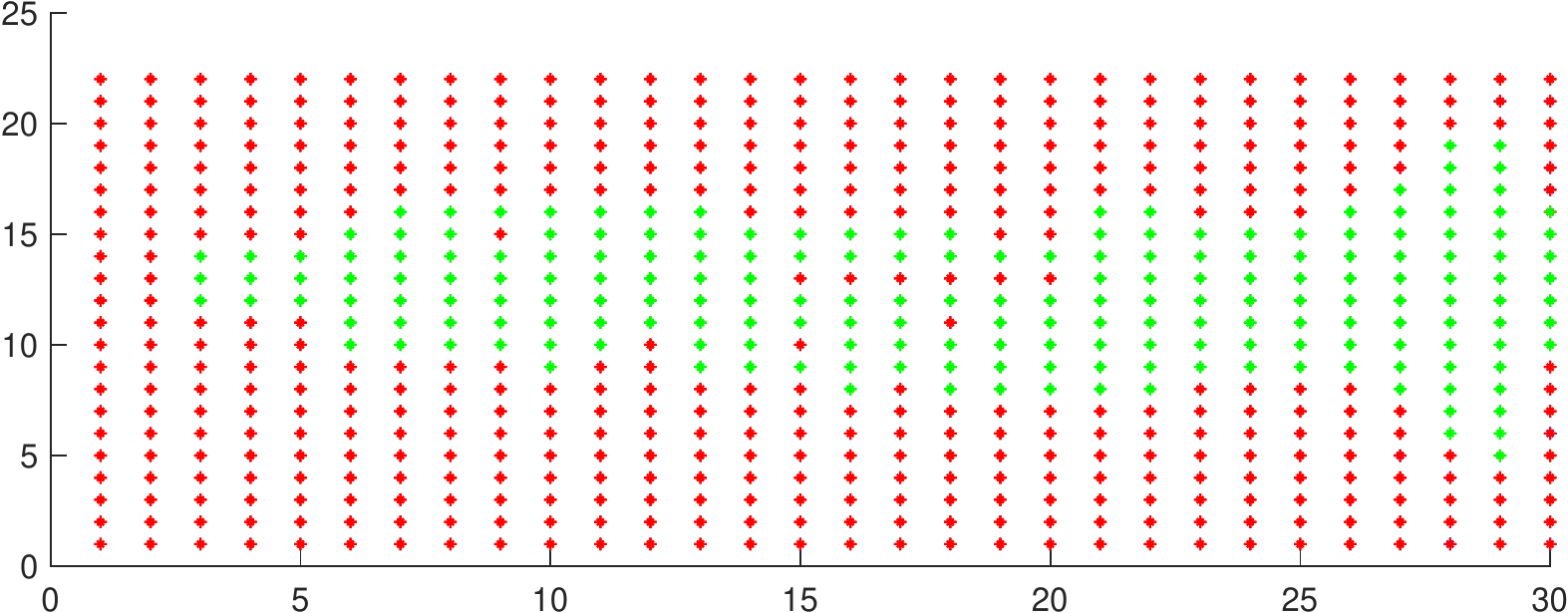}
    \caption{\small \emph{D$^2$-CAMP} on Sculpture dataset at time point 9}
    \label{fig:specalgo_s30_9_s}
\end{subfigure}

\begin{subfigure}[ht!]{.5\columnwidth}
    \includegraphics[width=\columnwidth]{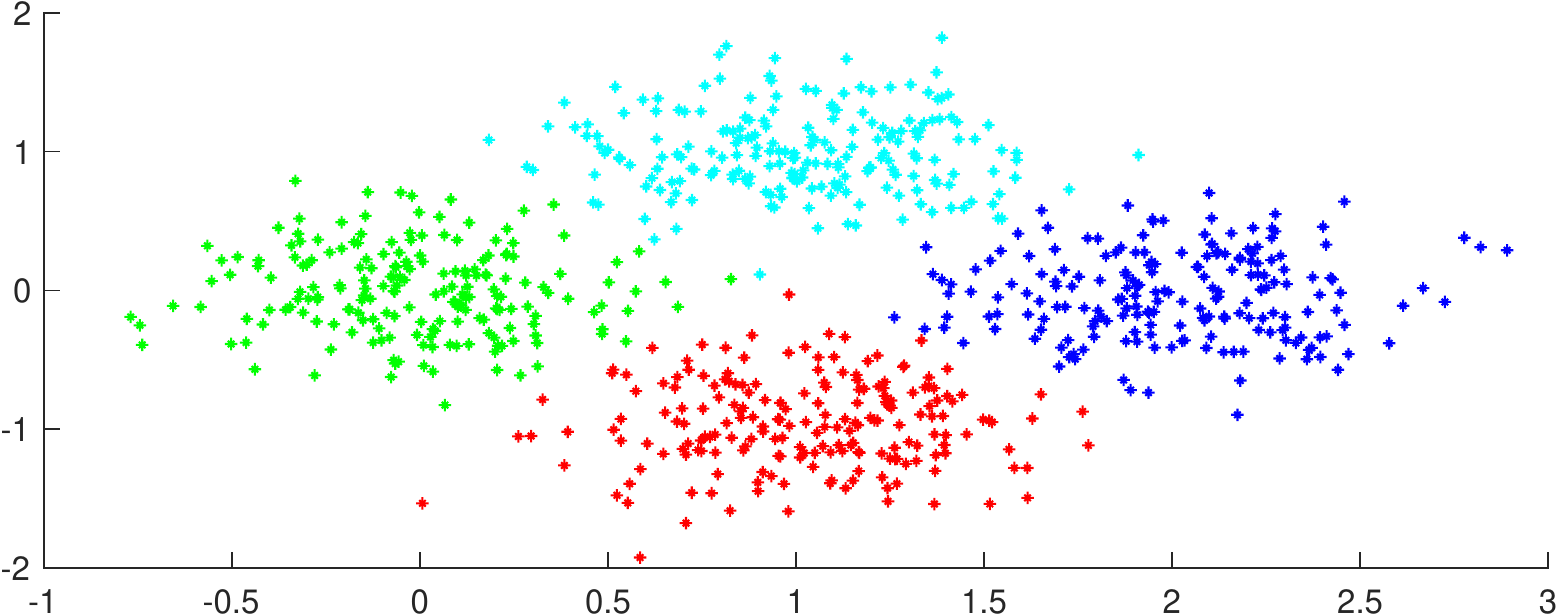}
    \caption{\small \emph{CNTRL} on Gaussians dataset at time point 10}
    \label{fig:baseline1_10_g}
\end{subfigure}
\begin{subfigure}[ht!]{.5\columnwidth}
    \includegraphics[width=\columnwidth]{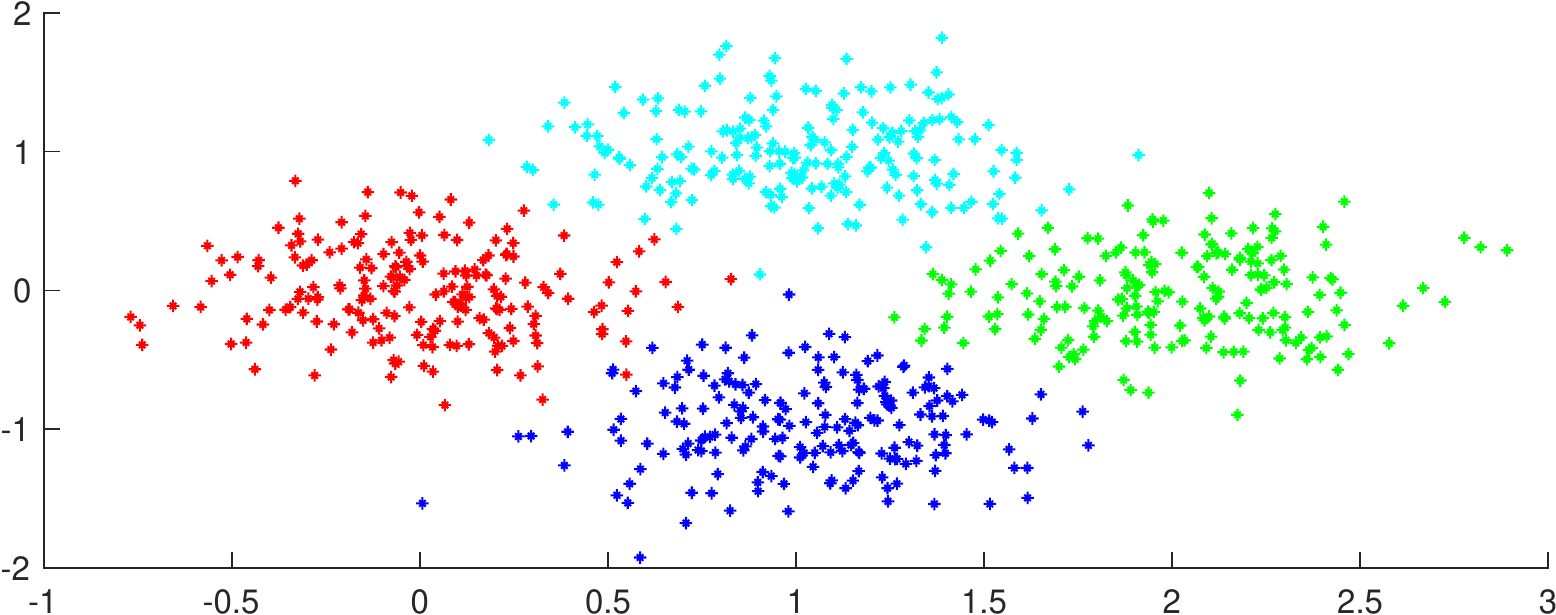}
    \caption{\small \emph{D$^2$-CAMP} on Gaussians dataset at time point 10}
    \label{fig:specalgo_s30_10_g}
\end{subfigure}
\begin{subfigure}[ht!]{.5\columnwidth}
    \includegraphics[width=\columnwidth]{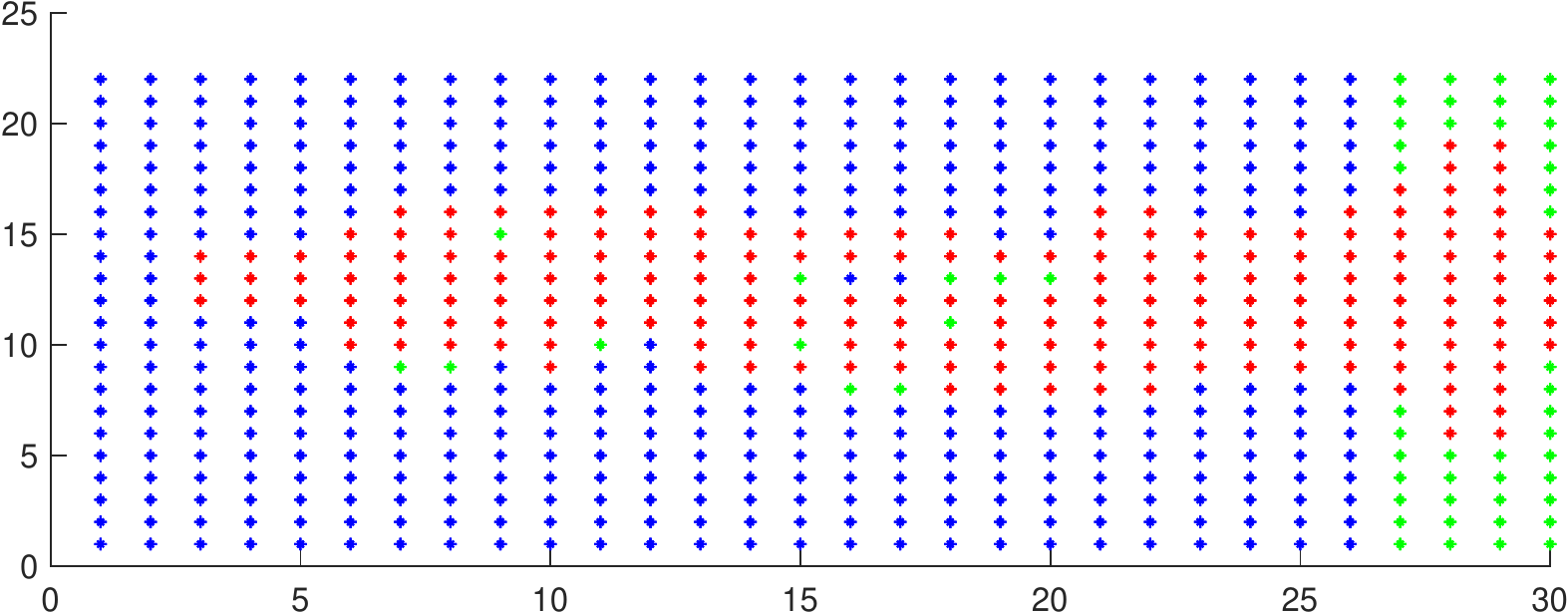}
    \caption{\small \emph{CNTRL} on Sculpture dataset at time point 10}
    \label{fig:baseline1_10_s}
\end{subfigure}
\begin{subfigure}[ht!]{.5\columnwidth}
    \includegraphics[width=\columnwidth]{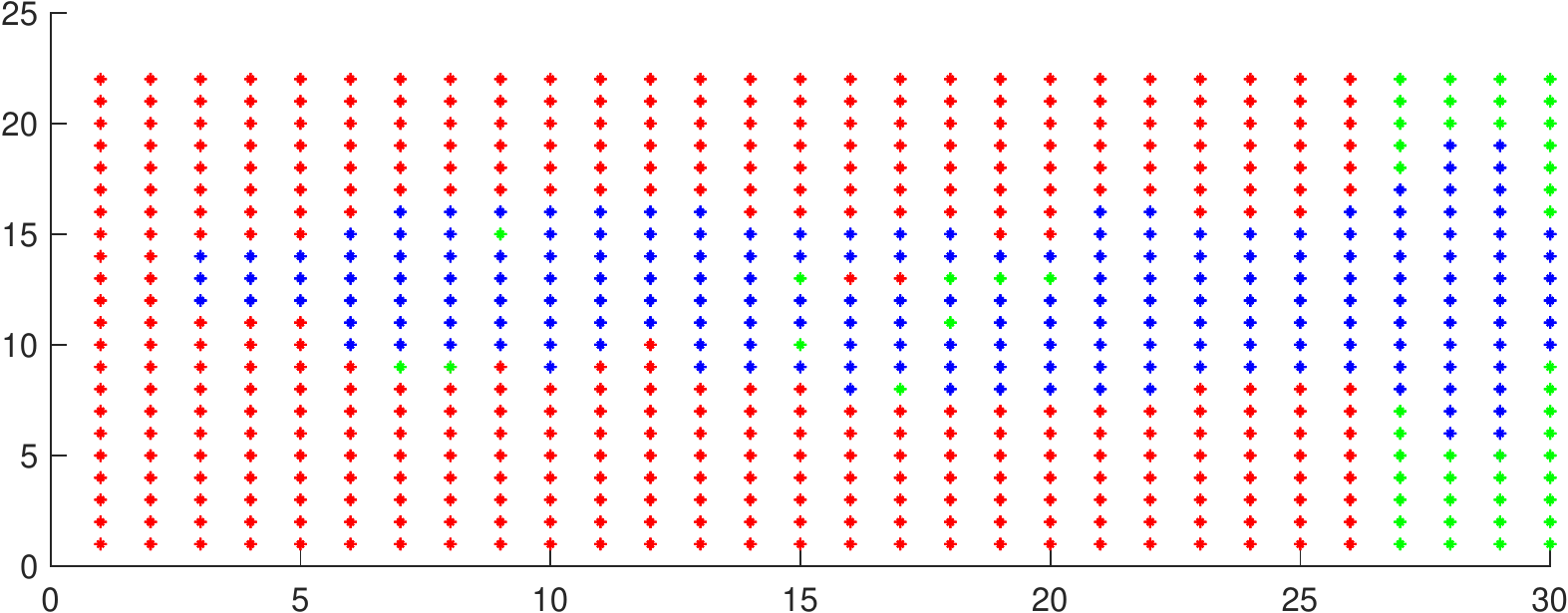}
    \caption{\small \emph{D$^2$-CAMP} on Sculpture dataset at time point 10}
    \label{fig:specalgo_s30_10_s}
\end{subfigure}

\caption{Communication cost, NCut and clustering results in the baseline setting}
\label{fig:baseline}
\vskip -0.2in
\end{figure*}

\begin{table}[t]
\center
\scriptsize
\begin{tabular}{l|c|c|c|c|c}
\hline
\multirow{2}{*}{Time} & \multirow{2}{*}{$s$} & Gaussians & Gaussians & Sculpture & Sculpture \\
& & \emph{D$^2$-CAMP} & \emph{D$^2$-CABL} & \emph{D$^2$-CAMP} & \emph{D$^2$-CABL} \\
\hline
\multirow{4}{*}{50}  & 15 & 4485 & 3132 & 15292 & 7130  \\
                     & 30 & 4607 & 3133& 15235 & 6054  \\
                     & 45 & 4660 & 3126& 15560 & 6076  \\
                     & 60 & 4669 & 3095& 15764 & 6705  \\
                     \hline
\multirow{4}{*}{90}  & 15 & 7342& 4988& 27036 & 12153\\
                     & 30 & 7533& 4982& 27020 & 10287 \\
                     & 45 & 7586& 4979& 27700 & 10336 \\
                     & 60 & 7630& 4960& 28001 & 11421 \\
                     \hline
\multirow{4}{*}{100} & 15 & 7748&5238& 28408 & 12846 \\
                     & 30 & 7988& 5230& 28338 & 10874 \\
                     & 45 & 7998& 5235& 29038 & 10897 \\
                     & 60 & 8062& 5218& 29343 & 12062\\
\hline
\end{tabular}
\caption{Communication cost with varied values of $s$}
\vspace{-0.1in}
\end{table}

\begin{table}[t]
\center
\scriptsize
\begin{tabular}{l|c|c|c|c|c}
\hline
\multirow{2}{*}{Time} & \multirow{2}{*}{$t$} & Gaussians & Gaussians & Sculpture & Sculpture \\
& & \emph{D$^2$-CAMP} & \emph{D$^2$-CABL} & \emph{D$^2$-CAMP} & \emph{D$^2$-CABL} \\\hline
\multirow{4}{*}{50\%}  & 10  & 4562 & 3127 & 15078 & 5998 \\
                       & 30  & 4645 & 3126 & 15278 & 6063 \\
                       & 100 & 4607 & 3133 & 15235 & 6054 \\
                       & 300 & 4620 & 3113 & 15269 & 6064 \\
                       \hline
\multirow{4}{*}{90\%}  & 10  & 7467 & 4979 & 26699 & 10202 \\
                       & 30  & 7581 & 4983 & 27012 & 10278 \\
                       & 100 & 7533 & 4982 & 27020 & 10287 \\
                       & 300 & 7618 & 4958 & 27042 & 10299 \\
                       \hline
\multirow{4}{*}{100\%} & 10  & 7917 & 5225 & 28046 & 10779 \\
                       & 30  & 8045 & 5234 & 28278 & 10847 \\
                       & 100 & 7988 & 5230 & 28338 & 10874 \\
                       & 300  & 8031 & 5211& 28345 & 10869\\
\hline
\end{tabular}
\caption{Communication cost with varied values of $t$}
\vspace{-0.15in}
\end{table}

\vspace{0.05in}
{\bf \noindent Experimental Results.}
As the baseline setting, we selected the total number of time points $t=10$ and the total number of sites $s=30$.
The communication cost and NCut of different algorithms on both datasets are shown in Fig. \ref{fig:baseline}.
On both datasets, the communication cost of \emph{D$^2$-CAMP} and \emph{D$^2$-CABL} are much smaller than \emph{CNTRL}, \emph{STMP} and \emph{STBL}.
Specifically, on Gaussians dataset, the communication cost of \emph{D$^2$-CAMP} can be only 4\% of that of \emph{STMP} and on average 16\% of that of \emph{CNTRL}.
The communication cost of \emph{D$^2$-CABL} is on average 11\% of \emph{CNTRL} and can be only 12\% of that of \emph{STBL}.
\emph{STMP} has communication cost even much larger than \emph{CNTRL}.
\emph{D$^2$-CABL} has a smaller communication cost than \emph{D$^2$-CAMP}.
On Sculpture dataset, the communication cost of \emph{D$^2$-CAMP} can be only 11\% of that of \emph{STMP} and is on average 49\% of that of \emph{CNTRL}.
The communication cost of \emph{D$^2$-CABL} can be only 15\% of that of \emph{STBL} and is on average 21\% of that of \emph{CNTRL}.
Similar to \emph{STMP}, \emph{STBL} also has communication cost larger than \emph{CNTRL}.
\emph{D$^2$-CABL} has a much smaller communication cost than \emph{D$^2$-CAMP} and the difference here is larger than in Gaussians dataset.

For both datasets, all algorithms have comparable NCut at every time point, except that on Gaussians dataset, at the time point 9, \emph{D$^2$-CABL} has a slightly larger NCut.
This could be due to that \emph{D$^2$-CABL} is a randomized algorithm with high success probability.
In Fig. \ref{fig:baseline}(e-l), the clustering results of \emph{CNTRL} and \emph{D$^2$-CAMP} on both datasets at time points 9 and 10 are visually very similar.
(The same cluster colors in different figures do not have relation.)
But for Sculpture dataset at the time point 9, the clustering result of \emph{D$^2$-CAMP} visually looks even more reasonable.

We then varied the value of $s$ from 15 to 60 with a step of 15 or the value of $t$ from 10 to 300 with a factor of 3 while keeping the other parameters unchanged as in the baseline setting.
Due to limit of space, we only show the resultant communication cost of \emph{D$^2$-CAMP} and \emph{D$^2$-CABL} on both datasets in Tables 1 and 2.
But the complete results are referred to Appendix.
When we varied the value of $s$, the communication cost of \emph{D$^2$-CAMP} increases roughly linearly with the increase of the value of $s$ from 15 to 60, while that of \emph{D$^2$-CABL} do not obviously increase with the value of $s$.
These observations are consistent with their theoretical communication cost $\tilde{O}(ns)$ and $\tilde{O}(n+s)$, respectively.
When we varied the value of $t$, both the communication cost of \emph{D$^2$-CAMP} and \emph{D$^2$-CABL} roughly keep the same, also supporting our theory above.

Finally, we tested the performance of \emph{D$^2$-CAMP} and \emph{D$^2$-CABL} for dynamic graph streams.
We randomly chose 5\% of edges to delete at a random time point after their arrival.
This increases the communicate cost of \emph{CNTRL} by 5\% as \emph{CNTRL} sends every deletion to the coordinator/blackboard.
However, the communication cost of \emph{D$^2$-CAMP} and \emph{D$^2$-CABL} are not changed.
More importantly, even ignoring the deletions, the resultant clusterings of \emph{D$^2$-CAMP} and \emph{D$^2$-CABL} at every time point have NCut comparable to that of \emph{CNTRL}.
Due to limit of space, we refer to Fig. 1 in Appendix.


\section{Conclusion and Future Work}

In this paper, we study the problem of how to efficiently perform graph clustering over modern graph data that are often \emph{dynamic} and collected at \emph{distributed} sites.
We design communication-optimal algorithms \emph{D$^2$-CAMP} and \emph{D$^2$-CABL} for two different communication models and prove their optimality rigorously.
Finally, we conducted extensive simulations to confirm that \emph{D$^2$-CAMP} and \emph{D$^2$-CABL} significantly outperform baseline algorithms in practice.
As the future work, we will study whether and how we can achieve similar results for geometric clustering, and how to achieve better computational bounds for the studied problems.
We will also study other related problems in the distributed dynamic setting such as low-rank approximation \cite{BKW17}, source-wise and standard round-trip spanner constructions \cite{ZL17,ZL18} and cut sparsifier constructions \cite{ADK+16}.

\section*{Acknowledgments}
This work was partially supported by NSF grants DBI- 1356655, CCF-1514357, IIS-1718738, as well as NIH grants R01DA037349 and K02DA043063 to Jinbo Bi.

\bibliography{cdgc}
\bibliographystyle{aaai}

\newpage
\section*{Appendix}
\setcounter{section}{0}
\setcounter{figure}{0}
\setcounter{table}{0}
\section{The Complete Results when varying the value of $s$ and $t$}
The results of communication cost and normalized cut (NCut) at every time point $\tau\in[1,t]$ when varying the value of $s$ on the Gaussians dataset and the Sculpture dataset are presented in Tables 1 and 2, respectively.
Basically, the communication cost increases linearly with respect to $s$ for \emph{D$^2$-CAMP}. The increase for \emph{D$^2$-CABL} are not obvious.
The results of communication cost and normalized cut (NCut) at every time point that is a multiple of 10\% of the total number of time points when varying the value of $t$ on the Gaussians dataset and the Sculpture dataset are presented in Tables 3 and 4, respectively.
The communication costs roughly keep unchanged for \emph{D$^2$-CAMP} and \emph{D$^2$-CABL}.
In all the tables, the NCut for different algorithms are comparably, except some rare cases when any algorithm do not succeed.
\begin{table*}[t]
\center
\caption{NCut and communication cost with varied values of $s$ on Gaussians dataset}
\begin{tabular}{l|c|c|c|c|c|c|r}
\hline
\multirow{2}{*}{Time} & \multirow{2}{*}{$s$} & \emph{CNTRL} & \emph{CNTRL} & \emph{D$^2$-CAMP} & \emph{D$^2$-CAMP} & \emph{D$^2$-CABL} & \emph{D$^2$-CABL} \\
& & NCut & Comm. & NCut & Comm. & NCut & Comm. \\
\hline
\multirow{4}{*}{10}  & 15 & \multirow{4}{*}{2.941} & \multirow{4}{*}{6556}  & 2.862 & 1025 & 2.953 & 784  \\
                     & 30 & &  & 2.921 & 1067 & 2.839 & 784  \\
                     & 45 & &  & 2.869 & 1083 & 2.913 & 772  \\
                     & 60 & &  & 2.96  & 1050 & 2.843 & 791  \\ \hline
\multirow{4}{*}{20}  & 15 & \multirow{4}{*}{2.918} & \multirow{4}{*}{11265} & 2.954 & 1784 & 2.923 & 1275 \\
                     & 30 & & & 2.954 & 1886 & 2.88  & 1284 \\
                     & 45 & & & 2.918 & 1916 & 2.881 & 1281 \\
                     & 60 & & & 2.982 & 1872 & 2.853 & 1263 \\ \hline
\multirow{4}{*}{30}  & 15 & \multirow{4}{*}{2.729} & \multirow{4}{*}{15872} & 2.855 & 2533 & 2.931 & 1800 \\
                     & 30 & & & 2.842 & 2651 & 2.932 & 1804 \\
                     & 45 & & & 2.93  & 2707 & 2.787 & 1831 \\
                     & 60 & & & 2.896 & 2612 & 2.905 & 1802 \\ \hline
\multirow{4}{*}{40}  & 15 & \multirow{4}{*}{2.744} & \multirow{4}{*}{21802} & 2.91  & 3510 & 2.707 & 2510 \\
                     & 30 & & & 2.854 & 3643 & 2.939 & 2500 \\
                     & 45 & & & 2.961 & 3698 & 2.886 & 2505 \\
                     & 60 & & & 2.856 & 3632 & 2.822 & 2481 \\ \hline
\multirow{4}{*}{50}  & 15 & \multirow{4}{*}{2.721} & \multirow{4}{*}{27748} & 2.763 & 4485 & 2.87  & 3132 \\
                     & 30 & & & 2.897 & 4607 & 2.909 & 3133 \\
                     & 45 & & & 2.748 & 4660 & 2.758 & 3126 \\
                     & 60 & & & 2.753 & 4669 & 2.814 & 3095 \\ \hline
\multirow{4}{*}{60}  & 15 & \multirow{4}{*}{2.712} & \multirow{4}{*}{32649} & 2.841 & 5297 & 2.785 & 3623 \\
                     & 30 & & & 2.829 & 5407 & 2.704 & 3623 \\
                     & 45 & & & 2.829 & 5473 & 2.866 & 3602 \\
                     & 60 & & & 2.651 & 5497 & 2.914 & 3597 \\ \hline
\multirow{4}{*}{70}  & 15 & \multirow{4}{*}{2.846} & \multirow{4}{*}{35976} & 2.853 & 5863 & 2.908 & 3959 \\
                     & 30 & & & 2.868 & 6003 & 2.743 & 3972 \\
                     & 45 & & & 2.707 & 6020 & 2.681 & 3956 \\
                     & 60 & & & 2.855 & 6086 & 2.823 & 3911 \\ \hline
\multirow{4}{*}{80}  & 15 & \multirow{4}{*}{2.794} & \multirow{4}{*}{39445} & 2.847 & 6430 & 2.854 & 4372 \\
                     & 30 & & & 2.766 & 6592 & 2.932 & 4377 \\
                     & 45 & & & 2.804 & 6599 & 2.844 & 4346 \\
                     & 60 & & & 2.875 & 6645 & 2.881 & 4312 \\ \hline
\multirow{4}{*}{90}  & 15 & \multirow{4}{*}{0.198} & \multirow{4}{*}{45250} & 0.216 & 7342 & 0.206 & 4988 \\
                     & 30 & & & 0.199 & 7533 & 0.22  & 4982 \\
                     & 45 & & & 1.102 & 7586 & 0.224 & 4979 \\
                     & 60 & & & 0.205 & 7630 & 0.207 & 4960 \\ \hline
\multirow{4}{*}{100} & 15 & \multirow{4}{*}{0.198} & \multirow{4}{*}{47897} & 0.208 & 7748 & 0.206 & 5238 \\
                     & 30 & & & 0.2   & 7988 & 0.22  & 5230 \\
                     & 45 & & & 0.206 & 7998 & 0.215 & 5235 \\
                     & 60 & & & 0.205 & 8062 & 0.204 & 5218\\
\hline
\end{tabular}
\end{table*}
\begin{table*}[t]
\center
\caption{NCut and communication cost with varied values of $s$ on Sculpture dataset}
\begin{tabular}{l|c|c|c|c|c|c|r}
\hline
\multirow{2}{*}{Time} & \multirow{2}{*}{$s$} & \emph{CNTRL} & \emph{CNTRL} & \emph{D$^2$-CAMP} & \emph{D$^2$-CAMP} & \emph{D$^2$-CABL} & \emph{D$^2$-CABL} \\
& & NCut & Comm. & NCut & Comm. & NCut & Comm. \\
\hline
\multirow{4}{*}{10}  & 15 & \multirow{4}{*}{1.874} & \multirow{4}{*}{5798}  & 1.991 & 3210  & 1.973 & 1574  \\
                     & 30 & &  & 1.121 & 3145  & 1.984 & 1348  \\
                     & 45 & &  & 1.12  & 3254  & 1.883 & 1344  \\
                     & 60 & &  & 1.117 & 3263  & 1.963 & 1491  \\ \hline
\multirow{4}{*}{20}  & 15 & \multirow{4}{*}{1.924} & \multirow{4}{*}{11792} & 1.971 & 6264  & 1.974 & 2928  \\
                     & 30 & & & 0.466 & 6210  & 1.947 & 2507  \\
                     & 45 & & & 0.475 & 6394  & 1.935 & 2511  \\
                     & 60 & & & 1.102 & 6447  & 1.817 & 2785  \\ \hline
\multirow{4}{*}{30}  & 15 & \multirow{4}{*}{1.698} & \multirow{4}{*}{18810} & 1.933 & 9503  & 1.846 & 4478  \\
                     & 30 & & & 1.087 & 9421  & 1.843 & 3834  \\
                     & 45 & & & 1.034 & 9659  & 1.988 & 3812  \\
                     & 60 & & & 1.091 & 9787  & 1.955 & 4241  \\ \hline
\multirow{4}{*}{40}  & 15 & \multirow{4}{*}{1.89}  & \multirow{4}{*}{25388} & 1.845 & 12562 & 1.97  & 5856  \\
                     & 30 & & & 0.434 & 12501 & 1.852 & 5019  \\
                     & 45 & & & 0.235 & 12798 & 1.806 & 4982  \\
                     & 60 & & & 0.23  & 12976 & 2.002 & 5546  \\ \hline
\multirow{4}{*}{50}  & 15 & \multirow{4}{*}{1.927} & \multirow{4}{*}{31256} & 1.765 & 15292 & 1.804 & 7130  \\
                     & 30 & & & 0.305 & 15235 & 1.788 & 6054  \\
                     & 45 & & & 0.653 & 15560 & 1.678 & 6076  \\
                     & 60 & & & 0.755 & 15764 & 1.965 & 6705  \\ \hline
\multirow{4}{*}{60}  & 15 & \multirow{4}{*}{1.742} & \multirow{4}{*}{37954} & 1.745 & 18434 & 1.929 & 8500  \\
                     & 30 & & & 1.079 & 18387 & 1.924 & 7233  \\
                     & 45 & & & 1.983 & 18798 & 1.889 & 7234  \\
                     & 60 & & & 1.043 & 19033 & 1.941 & 7997  \\ \hline
\multirow{4}{*}{70}  & 15 & \multirow{4}{*}{1.949} & \multirow{4}{*}{44566} & 1.823 & 21436 & 1.952 & 9877  \\
                     & 30 & & & 1.948 & 21421 & 1.726 & 8378  \\
                     & 45 & & & 1.835 & 21888 & 1.911 & 8394  \\
                     & 60 & & & 1.39  & 22156 & 1.939 & 9264  \\ \hline
\multirow{4}{*}{80}  & 15 & \multirow{4}{*}{1.892} & \multirow{4}{*}{51437} & 0.086 & 24676 & 1.914 & 11329 \\
                     & 30 & & & 1.856 & 24654 & 1.845 & 9598  \\
                     & 45 & & & 1.56  & 25225 & 1.867 & 9633  \\
                     & 60 & & & 1.848 & 25512 & 1.726 & 10647 \\ \hline
\multirow{4}{*}{90}  & 15 & \multirow{4}{*}{1.945} & \multirow{4}{*}{56331} & 1.749 & 27036 & 1.779 & 12153 \\
                     & 30 & & & 1.878 & 27020 & 1.825 & 10287 \\
                     & 45 & & & 1.695 & 27700 & 1.906 & 10336 \\
                     & 60 & & & 1.868 & 28001 & 1.774 & 11421 \\ \hline
\multirow{4}{*}{100} & 15 & \multirow{4}{*}{0.009} & \multirow{4}{*}{61452} & 0.01  & 28408 & 0.011 & 12846 \\
                     & 30 & & & 0.009 & 28338 & 0.011 & 10874 \\
                     & 45 & & & 0.009 & 29038 & 0.009 & 10897 \\
                     & 60 & & & 0.013 & 29343 & 0.013 & 12062\\
\hline
\end{tabular}
\end{table*}
\begin{table*}[t]
\center
\caption{NCut and communication cost with varied values of $t$ on Gaussians dataset}
\begin{tabular}{l|c|c|c|c|c|c|r}
\hline
\multirow{2}{*}{Time} & \multirow{2}{*}{$t$} & \emph{CNTRL} & \emph{CNTRL} & \emph{D$^2$-CAMP} & \emph{D$^2$-CAMP} & \emph{D$^2$-CABL} & \emph{D$^2$-CABL} \\
& & NCut & Comm. & NCut & Comm. & NCut & Comm. \\
\hline
\multirow{4}{*}{10\%}  & 10  &  \multirow{4}{*}{2.941} &  \multirow{4}{*}{6556}  & 2.869 & 1091 & 2.96  & 763  \\
                       & 30  & &  & 2.927 & 1148 & 2.882 & 790  \\
                       & 100 & &  & 2.921 & 1067 & 2.839 & 784  \\
                       & 300 & &  & 2.975 & 1152 & 2.806 & 725  \\ \hline
\multirow{4}{*}{20\%}  & 10  &  \multirow{4}{*}{2.918} &  \multirow{4}{*}{11265} & 2.965 & 1857 & 2.866 & 1253 \\
                       & 30  & & & 2.822 & 1935 & 2.908 & 1261 \\
                       & 100 & & & 2.954 & 1886 & 2.88  & 1284 \\
                       & 300 & & & 2.979 & 1929 & 2.863 & 1189 \\ \hline
\multirow{4}{*}{30\%}  & 10  &  \multirow{4}{*}{2.729} &  \multirow{4}{*}{15872} & 2.788 & 2569 & 2.981 & 1820 \\
                       & 30  & & & 2.868 & 2692 & 2.895 & 1816 \\
                       & 100 & & & 2.842 & 2651 & 2.932 & 1804 \\
                       & 300 & & & 2.926 & 2700 & 2.868 & 1709 \\ \hline
\multirow{4}{*}{40\%}  & 10  &  \multirow{4}{*}{2.744} &  \multirow{4}{*}{21802} & 2.91  & 3568 & 2.692 & 2521 \\
                       & 30  & & & 2.832 & 3680 & 2.891 & 2516 \\
                       & 100 & & & 2.854 & 3643 & 2.939 & 2500 \\
                       & 300 & & & 2.834 & 3648 & 2.844 & 2438 \\ \hline
\multirow{4}{*}{50\%}  & 10  &  \multirow{4}{*}{2.721} &  \multirow{4}{*}{27748} & 2.945 & 4562 & 2.771 & 3127 \\
                       & 30  & & & 2.797 & 4645 & 2.889 & 3126 \\
                       & 100 & & & 2.897 & 4607 & 2.909 & 3133 \\
                       & 300 & & & 2.883 & 4620 & 2.846 & 3113 \\ \hline
\multirow{4}{*}{60\%}  & 10  &  \multirow{4}{*}{2.712} &  \multirow{4}{*}{32649} & 2.904 & 5397 & 2.749 & 3616 \\
                       & 30  & & & 2.861 & 5479 & 2.807 & 3616 \\
                       & 100 & & & 2.829 & 5407 & 2.704 & 3623 \\
                       & 300 & & & 2.706 & 5465 & 2.689 & 3599 \\ \hline
\multirow{4}{*}{70\%}  & 10  & \multirow{4}{*}{2.846} & \multirow{4}{*}{35976} & 2.821 & 5971 & 2.944 & 3948 \\
                       & 30  & & & 2.855 & 6070 & 2.814 & 3953 \\
                       & 100 & & & 2.868 & 6003 & 2.743 & 3972 \\
                       & 300 & & & 2.825 & 6044 & 2.905 & 3913 \\ \hline
\multirow{4}{*}{80\%}  & 10  &  \multirow{4}{*}{2.794} &  \multirow{4}{*}{39445} & 2.749 & 6538 & 2.851 & 4348 \\
                       & 30  & & & 2.876 & 6650 & 2.816 & 4346 \\
                       & 100 & & & 2.766 & 6592 & 2.932 & 4377 \\
                       & 300 & & & 2.829 & 6616 & 2.809 & 4316 \\ \hline
\multirow{4}{*}{90\%}  & 10  &  \multirow{4}{*}{0.198} &  \multirow{4}{*}{45250} & 0.209 & 7467 & 1.042 & 4979 \\
                       & 30  & & & 1.033 & 7581 & 0.221 & 4983 \\
                       & 100 & & & 0.199 & 7533 & 0.22  & 4982 \\
                       & 300 & & & 1.039 & 7618 & 1.017 & 4958 \\ \hline
\multirow{4}{*}{100\%} & 10  &  \multirow{4}{*}{0.198} &  \multirow{4}{*}{47897} & 0.205 & 7917 & 0.206 & 5225 \\
                       & 30  & & & 0.204 & 8045 & 0.215 & 5234 \\
                       & 100 & & & 0.2   & 7988 & 0.22  & 5230 \\
                       & 300 & & & 0.202 & 8031 & 0.211 & 5211\\
\hline
\end{tabular}
\end{table*}
\begin{table*}[t]
\center
\caption{NCut and communication cost with varied values of $t$ on Sculpture dataset}
\begin{tabular}{l|c|c|c|c|c|c|r}
\hline
\multirow{2}{*}{Time} & \multirow{2}{*}{$t$} & \emph{CNTRL} & \emph{CNTRL} & \emph{D$^2$-CAMP} & \emph{D$^2$-CAMP} & \emph{D$^2$-CABL} & \emph{D$^2$-CABL} \\
& & NCut & Comm. & NCut & Comm. & NCut & Comm. \\
\hline
\multirow{4}{*}{10\%}  & 10  & \multirow{4}{*}{1.874} & \multirow{4}{*}{5798} & 1.106 & 3207  & 1.988 & 1361  \\
                       & 30  & &  & 1.121 & 3204  & 1.976 & 1341  \\
                       & 100 & &  & 1.121 & 3145  & 1.984 & 1348  \\
                       & 300 & &  & 1.115 & 3276  & 1.974 & 1360  \\ \hline
\multirow{4}{*}{20\%}  & 10  & \multirow{4}{*}{1.924} & \multirow{4}{*}{11792} & 1.092 & 6188  & 1.967 & 2486  \\
                       & 30  & & & 0.463 & 6277  & 1.994 & 2540  \\
                       & 100 & & & 0.466 & 6210  & 1.947 & 2507  \\
                       & 300 & & & 1.088 & 6296  & 1.945 & 2497  \\ \hline
\multirow{4}{*}{30\%}  & 10  & \multirow{4}{*}{1.698} & \multirow{4}{*}{18810} & 0.283 & 9399  & 1.813 & 3804  \\
                       & 30  & & & 0.977 & 9489  & 1.75  & 3865  \\
                       & 100 & & & 1.087 & 9421  & 1.843 & 3834  \\
                       & 300 & & & 1.104 & 9505  & 1.902 & 3861  \\ \hline
\multirow{4}{*}{40\%}  & 10  & \multirow{4}{*}{1.89}  & \multirow{4}{*}{25388} & 0.232 & 12407 & 1.937 & 4952  \\
                       & 30  & & & 0.571 & 12593 & 1.884 & 5045  \\
                       & 100 & & & 0.434 & 12501 & 1.852 & 5019  \\
                       & 300 & & & 0.239 & 12573 & 1.982 & 4997  \\ \hline
\multirow{4}{*}{50\%}  & 10  & \multirow{4}{*}{1.927} & \multirow{4}{*}{31256} & 0.448 & 15078 & 1.848 & 5998  \\
                       & 30  & & & 0.2   & 15278 & 1.967 & 6063  \\
                       & 100 & & & 0.305 & 15235 & 1.788 & 6054  \\
                       & 300 & & & 0.562 & 15269 & 1.69  & 6064  \\ \hline
\multirow{4}{*}{60\%}  & 10  & \multirow{4}{*}{1.742} & \multirow{4}{*}{37954} & 1.079 & 18195 & 1.924 & 7139  \\
                       & 30  & & & 0.93  & 18464 & 1.784 & 7201  \\
                       & 100 & & & 1.079 & 18387 & 1.924 & 7233  \\
                       & 300 & & & 0.952 & 18392 & 1.8   & 7221  \\ \hline
\multirow{4}{*}{70\%}  & 10  & \multirow{4}{*}{1.949} & \multirow{4}{*}{44566} & 1.942 & 21171 & 1.927 & 8289  \\
                       & 30  & & & 1.957 & 21400 & 1.938 & 8330  \\
                       & 100 & & & 1.948 & 21421 & 1.726 & 8378  \\
                       & 300 & & & 1.934 & 21387 & 1.865 & 8377  \\ \hline
\multirow{4}{*}{80\%}  & 10  & \multirow{4}{*}{1.892} & \multirow{4}{*}{51437} & 1.904 & 24363 & 2.001 & 9496  \\
                       & 30  & & & 1.73  & 24582 & 1.901 & 9566  \\
                       & 100 & & & 1.856 & 24654 & 1.845 & 9598  \\
                       & 300 & & & 1.839 & 24647 & 1.868 & 9609  \\ \hline
\multirow{4}{*}{90\%}  & 10  & \multirow{4}{*}{1.945} & \multirow{4}{*}{56331} & 1.902 & 26699 & 1.845 & 10202 \\
                       & 30  & & & 1.941 & 27012 & 1.879 & 10278 \\
                       & 100 & & & 1.878 & 27020 & 1.825 & 10287 \\
                       & 300 & & & 1.911 & 27042 & 1.755 & 10299 \\ \hline
\multirow{4}{*}{100\%} & 10  & \multirow{4}{*}{0.009} & \multirow{4}{*}{61452} & 0.011 & 28046 & 0.011 & 10779 \\
                       & 30  & & & 0.011 & 28278 & 0.012 & 10847 \\
                       & 100 & & & 0.009 & 28338 & 0.011 & 10874 \\
                       & 300 & & & 0.01  & 28345 & 0.013 & 10869\\
\hline
\end{tabular}
\end{table*}

\section{The Complete Results for Dynamic Graph Streams}
The complete results of communication cost and NCut under dynamic graph stream on the Gaussians and Sculpture datasets are plotted in Fig. 1.
It can be seen that even though \emph{D$^2$-CAMP} and \emph{D$^2$-CABL} do not process the deletions, their NCut remains comparable to that of \emph{CNTRL}.
The communication cost can be saved by this trick, keeping much smaller than the communication cost of \emph{CNTRL}.
\begin{figure*}[ht]
\center
\begin{subfigure}[ht!]{.8\columnwidth}
     \centerline{\includegraphics[width=\columnwidth]{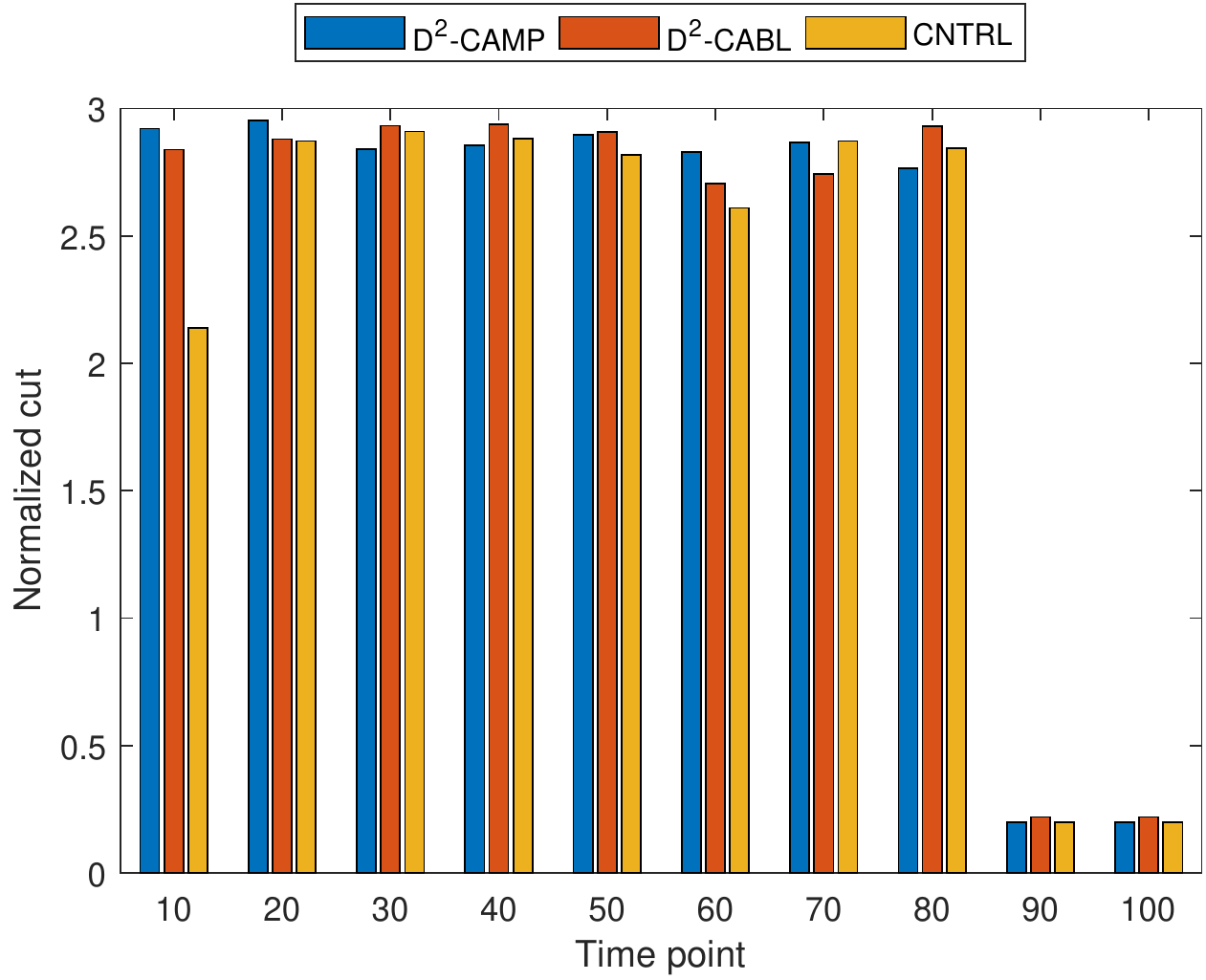}}
     \caption{\small NCut on Gaussians dataset}
     \label{fig:gncdynamic}
\end{subfigure}
\begin{subfigure}[ht!]{.8\columnwidth}
     \centerline{\includegraphics[width=\columnwidth]{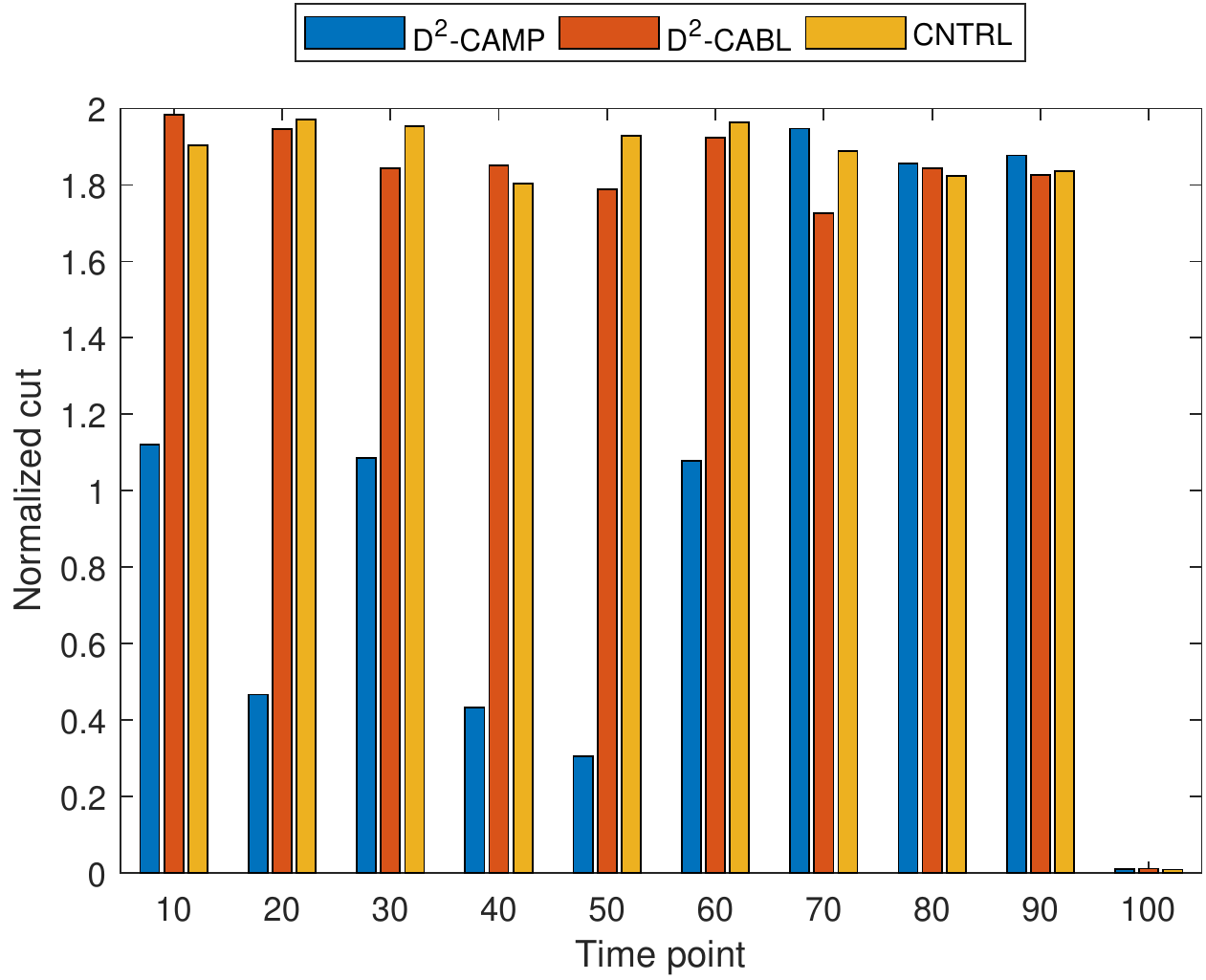}}
     \caption{\small NCut on Sculpture dataset}
     \label{fig:sncdynamic}
\end{subfigure}
\begin{subfigure}[ht!]{.8\columnwidth}
     \centerline{\includegraphics[width=\columnwidth]{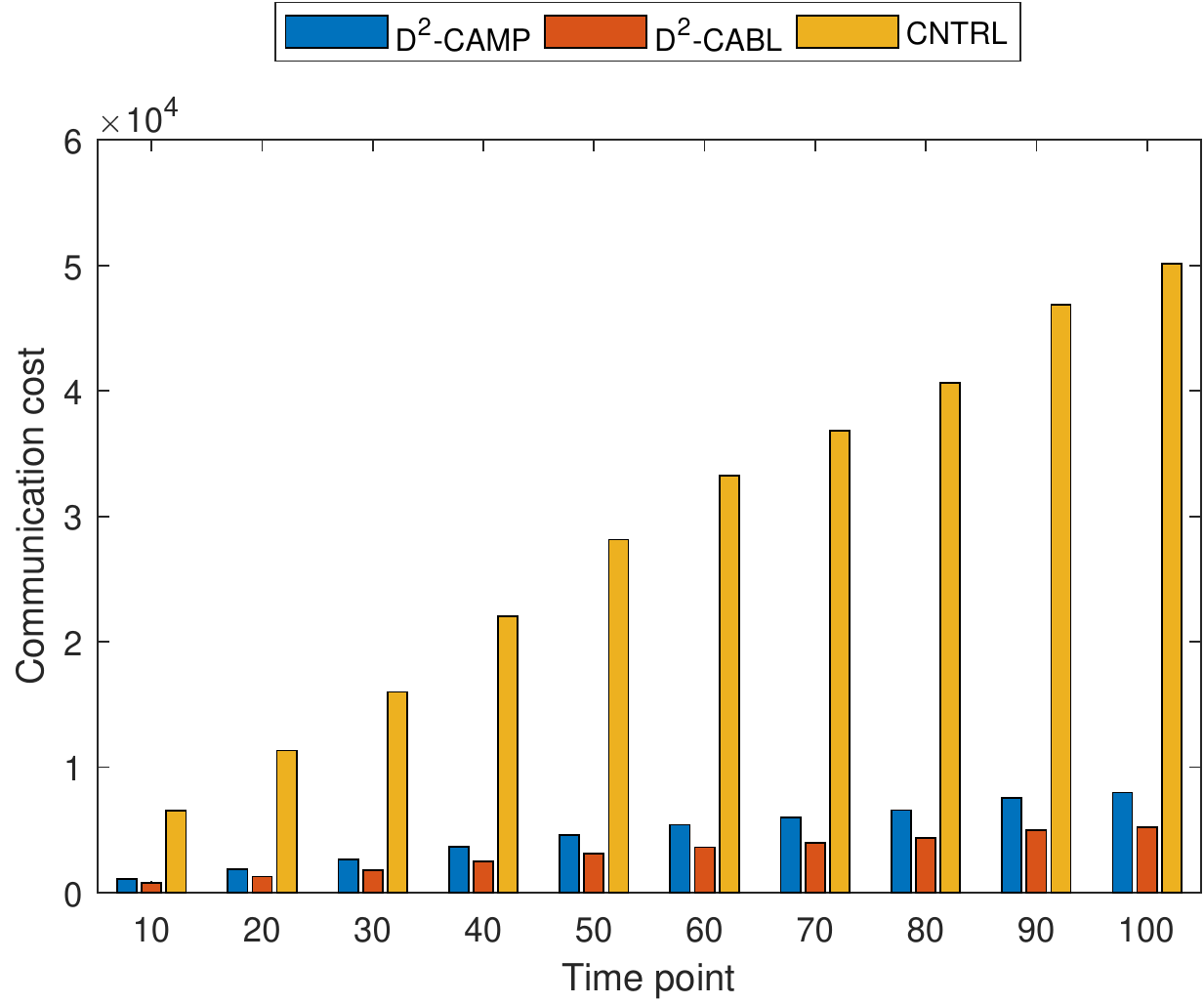}}
     \caption{\small Communication cost on Gaussians dataset}
     \label{fig:gccdynamic}
\end{subfigure}
\begin{subfigure}[ht!]{.8\columnwidth}
     \centerline{\includegraphics[width=\columnwidth]{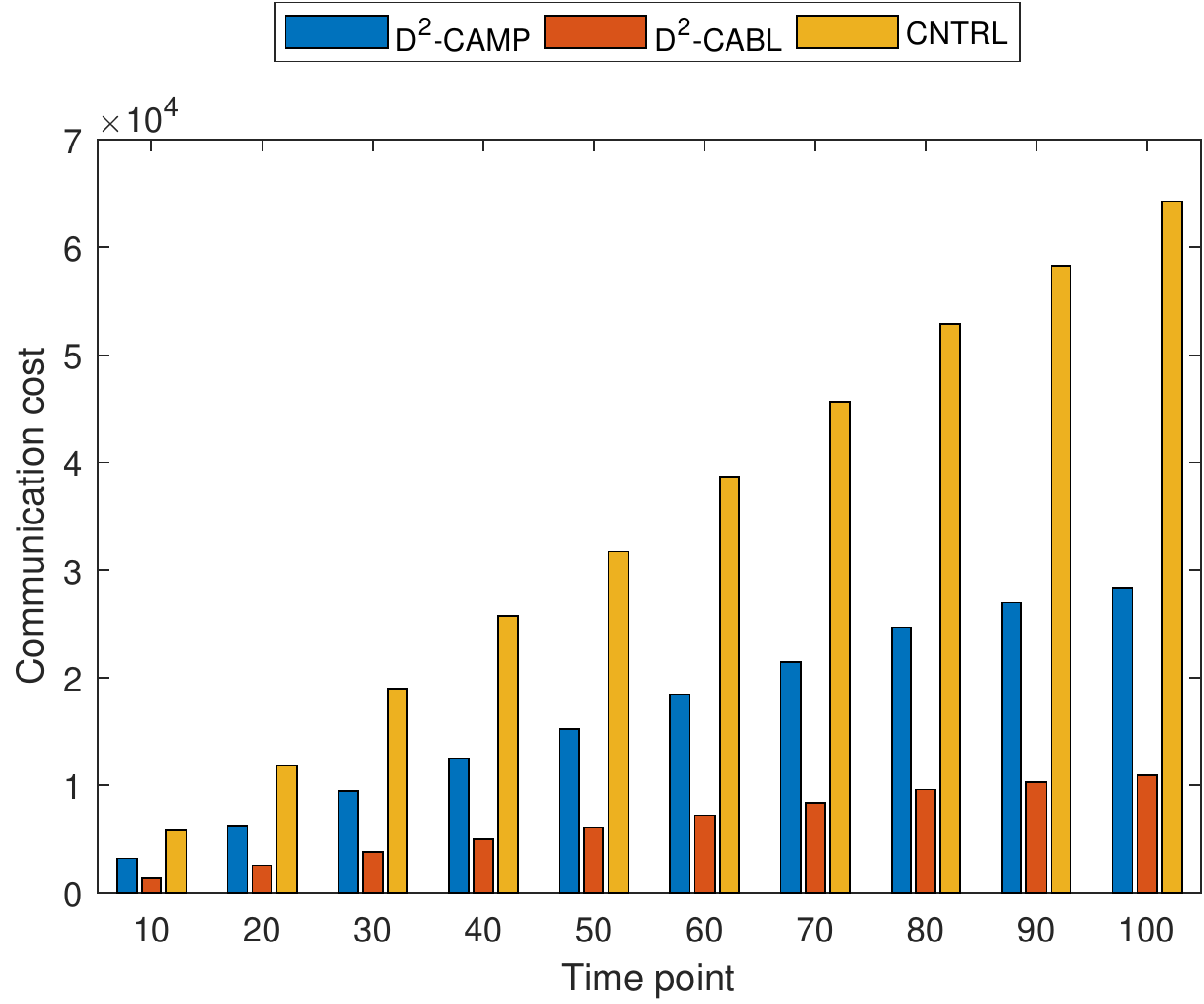}}
     \caption{\small Communication cost on Sculpture dataset}
     \label{fig:sccdynamic}
\end{subfigure}

\caption{The Complete Results for Dynamic Graph Streams ($t=100$ and $s=30$)}
\label{fig:baseline}
\vskip -0.2in
\end{figure*}

\section{Proof of Theorem 5}
\addtocounter{theorem}{-1}
\begin{theorem}
\label{thm:spanner}
Given two nodes $u,v\in V$ and an integer $k>1$, for every time point $\tau\in [1,t]$, the proposed algorithm can answer approximate shortest distance between $u$ and $v$ in $G^{\tau}$ no larger than $2k-1$ times of their actual shortest distance at the coordinator in the message passing model.
Summing over $t$ time points, the total communication cost is $\tilde{O}(n^{1+1/k}s)$.
\end{theorem}
\proof
We first prove that at every time $\tau$, the constructed subgraph $Q^{\tau}=\cup_{i=1}^iQ^{\tau}_i$ is a $(2k-1)$-spanner of the graph $G^{\tau}=\cup_{i=1}^iG^{\tau}_i$ received up to the time point $\tau$.
For each edge $e=(u,v)\in E^{\tau}_i$, there is a path $P$ between $u$ and $v$ in the spanner $Q^{\tau}_i$ of distance no larger than $(2k-1)W(e)$, because $Q^{\tau}_i$ is a $(2k-1)$-spanner of $G^{\tau}_i(V,E^{\tau}_i)$.
Then in the union graph $Q^{\tau}=\cup_{i=1}^iQ^{\tau}_i$, the path $P$ is still presented.
Therefore, for every edge $e(u,v)$ in $G^{\tau}$, there is a path between $u$ and $v$ in $Q^{\tau}$ with distance no larger than $(2k-1)W(e)$.
This implies that $Q^{\tau}$ is a $(2k-1)$-spanner of $G^{\tau}$.

By the monotonicity property, each site only needs to transmit $\tilde{O}(n^{1+1/k})$ summing over all $t$ time points.
Summing over $s$ sites, the total communication cost is $\tilde{O}(n^{1+1/k}s)$.
\qed

\end{document}